\documentclass[sigconf]{acmart} %
\usepackage{amsmath,amsfonts} %
\usepackage{algorithmic}
\usepackage{graphicx}
\usepackage{textcomp}
\usepackage{xcolor}
\usepackage{gensymb} %
\usepackage{xspace}
\usepackage{subcaption}
\usepackage{algorithm}
\usepackage{listings}

\definecolor{ForestGreen}{RGB}{34,139,34}

\newcommand{\DESIGN}{\mbox{Wattchmen}\xspace}

\copyrightyear{2026}
\acmYear{2026}
\setcopyright{usgovmixed}
\acmConference[ICS '26]{2026 International Conference on Supercomputing}{July 06--09, 2026}{Belfast, United Kingdom}
\acmBooktitle{2026 International Conference on Supercomputing (ICS '26), July 06--09, 2026, Belfast, United Kingdom}
\acmDOI{10.1145/3797905.3800514}
\acmISBN{979-8-4007-2522-7/2026/07}

\begin{document}

\title{Wattchmen: Watching the Wattchers -- High Fidelity, Flexible GPU Energy Modeling}
\author{Brandon Tran}
\affiliation{%
  \institution{University of Wisconsin-Madison}
  \city{Madison}
  \state{WI}
  \country{USA}
}
\email{bqtran2@wisc.edu}

\author{Matthias Maiterth}
\authornote{This work was done while this author was at Oak Ridge National Lab.}
\affiliation{%
  \institution{NVIDIA}
  \city{Santa Clara}
  \state{CA}
  \country{USA}}
\email{mmaiterth@nvidia.com}

\author{Woong Shin}
\affiliation{%
  \institution{Oak Ridge National Lab}
  \city{Oak Ridge}
  \state{TN}
  \country{USA}
}
\email{shinw@ornl.gov}

\author{Matthew D. Sinclair}
\affiliation{%
 \institution{University of Wisconsin-Madison}
 \city{Madison}
 \state{WI}
 \country{USA}}
\email{sinclair@cs.wisc.edu}

\author{Shivaram Venkataraman}
\affiliation{%
 \institution{University of Wisconsin-Madison}
 \city{Madison}
 \state{WI}
 \country{USA}}
\email{shivaram@cs.wisc.edu}

\begin{abstract}

Modern GPU-rich HPC systems are increasingly becoming energy-constrained.
Thus, understanding an application's energy consumption becomes essential. %
Unfortunately, current GPU energy attribution techniques are either inaccurate, inflexible, or outdated.
Therefore, we propose \DESIGN{}, a flexible methodology for measuring, attributing, and predicting GPU energy consumption.
We construct a per-instruction energy model using a diverse set of microbenchmarks to systematically quantify the energy consumption of GPU instructions, enabling finer-grain prediction and energy consumption breakdowns for applications.
Compared with the state-of-the-art systems like AccelWattch (32\%) and Guser (25\%), across 16 popular GPGPU, graph analytics, HPC, and ML workloads, \DESIGN{} reduces the mean absolute percent error (MAPE) to 14\% on V100 GPUs.
Furthermore, we show that \DESIGN{} provides similar MAPEs for water-cooled V100s (15\%) and extends to later architectures, including air-cooled A100 (11\%) and H100 (12\%) GPUs.
Finally, to further demonstrate \DESIGN{}'s value, we apply it to applications such as Backprop and QMCPACK, where \DESIGN{}'s insights enable energy reductions of up to 35\%.

\end{abstract}



\maketitle

\section{Introduction}
\label{sec:intro}

The slowing of Moore's Law %
introduced significant challenges for realizing exascale supercomputing~\cite{bergman:2008:exascalecomputing, shalf:2011}.
In turn, exascale class systems~\cite{top500} were realized by heavily relying on GPUs~\cite{Atchley2023}.
This trend continues: new high performance computing (HPC) systems use an ever larger number of modern GPUs. %
For example, the US Department of Energy's (DOE's) Aurora, El Capitan, and Frontier supercomputers have 37000$-$64000 GPUs. %
Although such systems are a major achievement, 
recent studies have shown that compute systems at all levels --- including full systems~\cite{Atchley2023, Kodama2020}, individual nodes, and GPUs~\cite{Choi2014, burger-isca2011, Sun2020, TschandRajan2025-mlperfPower} --- are \emph{power limited}~\cite{Maiterth2017, SolorzanoSato2024-fugakuPoints}.
Thus, moving forward, the HPC community will repeatedly hit the power wall, bound by the fundamental laws of technology scaling~\cite{powerwall, ShehabiNewkirk2024-doeDCEnergy}.  %
To overcome this barrier, researchers have identified four pillars to improve energy efficiency for HPC: site infrastructure, system hardware, system software, and applications~\cite{torsten_wilde_4_2014}.
While recent work has made significant progress in improving the first three pillars~\cite{ljungdhal2022-modelcooling, Ferdaus2025-evalaiaccelerator, wilson_end--end_2023},
we focus on the role of HPC applications in this paper.

Historically, 
HPC developers focused on optimizing their workloads for performance -- e.g., reaching exascale performance.
However, in modern systems, an application's compute performance is limited by its power draw.
Thus, developers must understand where this power budget is spent
to increase compute efficiency~\cite{AngChien2022-recode, Mudge2001-powerFirstClass, shin_2024_towards}.
To perform these informed optimizations, developers require credible tools that can accurately break down an application's energy consumption and attribute it to finer-grained components.
Traditionally, developers rely on modeling and simulation (ModSim) techniques to model and prototype energy consumption~\cite{Li2013-mcpat, cacti}.
Unfortunately, existing frameworks are \textbf{severely lacking} in helping developers and system designers analyze and break down the energy use of applications (discussed further in Section~\ref{sec:back}).

\begin{figure}[tb!]
    \centering
    \vspace{-1ex}
    \includegraphics[scale=0.22]{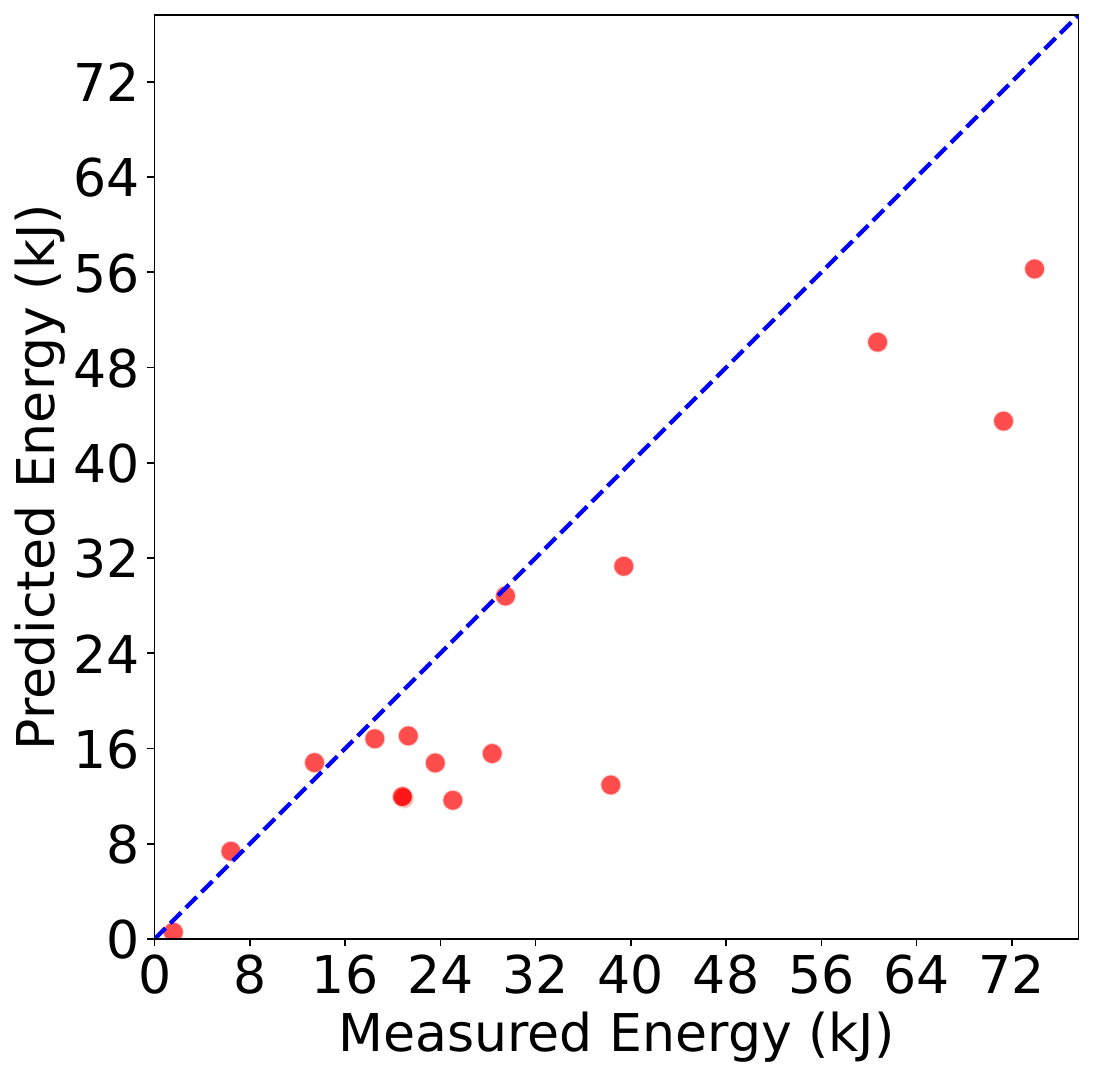}
    \vspace{-1ex}
    \caption{Comparing AccelWattch's energy predictions to measurements from air-cooled Tesla V100 GPUs across various benchmarks, which leads to a 32\% MAPE. The blue line indicates perfect prediction.}
    \Description{Comparing AccelWattch's energy predictions to measurements from air-cooled Tesla V100 GPUs across various benchmarks, which leads to a 32\% MAPE.}
    \label{fig:accelwattch_summit}
    \vspace{-2ex}
\end{figure}

\begin{table*}[tb!]
  \centering
  \caption{Comparing \DESIGN{} to recent GPU energy modeling approaches.}  
  {\scriptsize
\begin{tabular}{|l|c|c|c|c|c|c|}
  \hline
  \textbf{Feature} & Arafa, et al.\cite{ArafaElWazir2020-nmsu}& Guser\cite{shan2024-guser} & Delestrac, et al.\cite{DelestracMiquel2024-montpellier} & ML Power Models\cite{Alavani2023-gpuMLpred, WuGreathouse2015-amdGPUPowerML} & AccelWattch\cite{KandiahPeverelle2021-accelWattch} & \DESIGN{} \\ \hline
  Portable across vendor architecture & \color{ForestGreen}\checkmark & \color{ForestGreen}\checkmark &  \color{ForestGreen}\checkmark & \color{ForestGreen}\checkmark & \color{red}X & \color{ForestGreen}\checkmark \\ \hline
  Adapts to different cooling policies & \color{red}X & \color{ForestGreen}\checkmark & \color{ForestGreen}\checkmark & \color{ForestGreen}\checkmark & \color{red}X & \color{ForestGreen}\checkmark \\ \hline
  Models Compute Energy & \color{ForestGreen}\checkmark & \color{ForestGreen}\checkmark & \color{red}X & \color{red}X & \color{ForestGreen}\checkmark & \color{ForestGreen}\checkmark \\ \hline
  Models Control Flow Energy & \color{red}X & \color{red}X & \color{red}X & \color{red}X & \color{ForestGreen}\checkmark & \color{ForestGreen}\checkmark \\ \hline
  Models Memory Hierarchy Energy & \color{red}X & \color{ForestGreen}\checkmark & \color{ForestGreen}\checkmark & \color{red}X & \color{ForestGreen}\checkmark & \color{ForestGreen}\checkmark \\ \hline
  Fine-Grained Energy Breakdown & \color{ForestGreen}\checkmark & \color{red}X & \color{ForestGreen}\checkmark & \color{red}X & \color{ForestGreen}\checkmark & \color{ForestGreen}\checkmark \\ \hline
  Comprehensive Energy Measurements & \color{red}X & \color{ForestGreen}\checkmark & \color{red}X & \color{red}X & \color{ForestGreen}\checkmark & \color{ForestGreen}\checkmark \\ \hline
\end{tabular}
}

  \label{tab:relWork}
  \vspace{-2ex}
\end{table*}

For example, several modeling-based approaches have been developed to attribute energy use.
Arafa et al. and Delestrac et al. break down energy consumption at a finer granularity, e.g., individual GPU instructions~\cite{ArafaElWazir2020-nmsu, DelestracMiquel2024-montpellier}.
However, these works each examine only a subset of the instruction set, either compute instructions or the memory hierarchy.
Since these tools targeted only a subset of GPU instructions, they ignored any contributions from other components during their measurements.
An alternative is to employ ML-based models to predict an application's energy consumption~\cite{WuGreathouse2015-amdGPUPowerML,Alavani2023-gpuMLpred}.
However, this work lacks explainability when attributing the energy consumption to application kernels: while its kernel-level energy breakdown is helpful, it limits the user's ability to explain the impact of optimizations.

Alternatively, AccelWattch~\cite{KandiahPeverelle2021-accelWattch}, the state-of-the-art GPU power modeling framework, categorizes an application's energy consumption into architectural components (discussed further in Section~\ref{subsubsec:back-alt-accelwattch}).
However, AccelWattch's predictions are brittle and only work well if applied to the same validated GPU in the same environment. 
In Figure~\ref{fig:accelwattch_summit}, we used AccelWattch to predict energy consumption for popular applications~\cite{CheBoyer2009-rodinia, CheSheaffer2010-rodinia, Narang2016, Narang2017-deepBench} on an NVIDIA V100 GPU in NSF's CloudLab cluster~\cite{DuplyakinRicci2019-cloudlab}.
Although AccelWattch was validated for V100 GPUs, since CloudLab's environment was different, AccelWattch has a mean absolute percent error (MAPE) of 32\% -- twice AccelWattch's self-reported error.
In Section~\ref{subsubsec:back-alt-accelwattch} we discuss these experiments further, including why tuning AccelWattch was insufficient.

To address these issues, we propose \DESIGN{}, a GPU energy modeling tool and methodology that provides high-fidelity, fine-grained energy profiling of the GPU core compute, control flow, and memory hierarchy.
Moreover, \DESIGN{} adapts to configuration differences %
(e.g., air- and water-cooling) and vendor architecture variations (e.g., NVIDIA's Volta V100, Ampere A100, and Hopper H100 GPUs).
To create this model, we use an extensive suite of microbenchmarks designed to isolate the energy consumption of compute, control flow, and memory hierarchy instructions.
Second, we leverage modern profiling capabilities to gain a complete understanding of all the instructions executed on the GPU.
Key to our design is the construction of a system of energy equations that can be solved to determine instruction-level energies.
By solving the system of equations, we not only solve for instructions that are difficult to isolate on their own but also account for the impact of instructions that are part of the overhead when running other microbenchmarks.
Using this information, we develop a model that predicts instruction-level GPU energy consumption for any given application, thereby providing finer-grained energy information.

Overall, across 16 popular GPGPU, graph analytics, HPC, and ML workloads, 
\DESIGN{} reduces the mean absolute percent error (MAPE) to 13\% versus 32\% for the state-of-the-art AccelWattch and 25\% for Guser.
Moreover, unlike AccelWattch, \DESIGN{} adapts and accurately predicts energy on systems with different cooling mechanisms: %
providing 15\% MAPE on the Summit supercomputer's water-cooled V100 GPUs.
Furthermore, \DESIGN{} generalizes to more modern GPUs, yielding MAPEs of 11\% for A100 GPUs and 12\% for H100 GPUs.
Finally, we present two case studies on Backprop and QMCPack that demonstrate how \DESIGN{} can be used to improve energy efficiency (up to 35\%) %
on existing systems.
In these case studies, we show how \DESIGN{}'s energy breakdown %
enables developers to improve the energy consumption of their workloads.
Thus, \DESIGN{} provides high-fidelity and fine-grained insights into a GPU application's energy consumption.
We plan to open-source \DESIGN{} %
to help reproduction and extension, allowing the community to co-optimize their workloads.

\section{Background}
\label{sec:back}

\subsection{Vendor-based Tools for Energy Profiling}
\label{subsec:back-gpuMeasure}

GPU vendors have provided tools such as the NVIDIA Management Library (\texttt{NVML})~\cite{noauthor_dcgm_nodate, noauthor_nvidia_nodate} or AMD's \texttt{rocm-smi}~\cite{noauthor_rocm_nodate-1, noauthor_rocm_nodate} to measure the energy consumption of their devices.
These tools provide measurements for the entire GPU and are exposed at system-level by the system integrator~\cite{noauthor_redfish_nodate, noauthor_pm_nodate, Thaler:2020, openbmc_event} or community driven unified interfaces~\cite{noauthor_power_nodate, noauthor_variorum_nodate}.
Measurement data streams from these interfaces are plumbed through various data collection tools and data management systems~\cite{agelastos_lightweight_2014, bartolini_paving_2019, borghesi_examon-x_2023, eastep_global_2017, netti_dcdb_2020} and have been used for application power management~\cite{Etinski2012, Krzywaniak2020, patki_comparing_2019, Rountree2012, yue_reaper_2023}, power-aware job scheduling~\cite{maiterth_energy_2018}, data center level power management~\cite{wilson_end--end_2023}, HPC power signatures~\cite{shin_revealing_2021, govind_comparing_2023}, and the race to exascale~\cite{bergman:2008:exascalecomputing, Dongara-racetoexascale:2019, Kothe-USexascale:2019}.
However, these focus on addressing the overall system energy efficiency.

Prior work has also explored the relationship between application performance and energy consumption~\cite{choi_roofline_2013, choi_how_2013, ghane_power_2018}.
However, such ideas, methods, and techniques have not made it into major vendor profiling tools~\cite{noauthor_nsight_nodate, noauthor_user_nodate-1} or are limited in capabilities~\cite{noauthor_cray_nodate} due to the lack of suitable granularity in both time and space.
With microsecond-level GPU kernel activities happening in increasingly large, heterogeneous GPUs, fine-grained energy consumption attribution is exceedingly difficult.
Thus, while vendor-provided tools provide some valuable details, they currently do not provide detailed, instruction-level granularity information on GPU energy consumption. %
Thus, recent work (and \DESIGN{}) instead seeks alternatives that provide  developers' more fine-grained information.

\subsection{GPU Instructions}
\label{subsec:back-gpuInstr}

Modern GPU energy modeling frameworks frequently utilize microbenchmarks to isolate the energy behavior of specific instructions (Section~\ref{subsec:back-alt}).
For example, to estimate the energy consumption of a GPU vector addition, these frameworks use microbenchmarks that perform a large number of vector additions. %
To reduce the number of non-vector addition instructions in the program, these microbenchmarks typically use inlined assembly and other optimizations, such as loop unrolling.
Thus, the resultant microbenchmark often contains a large number of the assembly instructions in question, enabling the frameworks to isolate their energy consumption.
For NVIDIA GPUs, this assembly-level programming is typically done in CUDA's PTX virtual ISA~\cite{cuda_ptx_isa}, which acts as an intermediate ISA across all NVIDIA GPUs.
Subsequently, the assembler optimizes the PTX instructions into SASS.
Since SASS assembly instructions are specific to a given NVIDIA GPU, they are often not portable.
This two-stage compilation process allows NVIDIA to generate highly optimized code for a given GPU without guaranteeing backwards ISA compatibility.
However, it also makes it difficult to isolate the behavior of specific SASS instructions, since the assembler may generate different SASS instructions.
In Sections~\ref{subsec:des-nnSolve} and \ref{subsec:des-coverage} we discuss how \DESIGN{} overcomes this challenge.

\subsection{Alternative Energy Profiling Approaches}
\label{subsec:back-alt}

Building on Section~\ref{sec:intro}, Table~\ref{tab:relWork} compares \DESIGN{} to a number of other recent works that provide alternative approaches to measure GPU energy consumption.
To effectively co-optimize GPU applications for performance and energy on HPC systems, an energy model must (i) be \emph{generalizable} -- including across GPU architectures within the same hardware generation, across different generations, or adapting to deployment changes such as different cooling configurations; (ii) provide fine-grained energy information about compute (e.g., vector adds and multiplies), control flow (e.g., move, set predicates) and memory hierarchy (e.g., memory accesses to specialized memories like scratchpads and constant memory, as well as the data caches and main memory); and (iii) provide comprehensive energy breakdowns.
Existing works fall short in one or more dimensions. %

There are also several energy-aware GPU approaches (discussed further in Section~\ref{sec:relWork}).
However, for existing HPC systems where hardware changes are not possible, the most popular approaches either project empirical GPU energy measurements on existing systems~\cite{ArafaElWazir2020-nmsu, DelestracMiquel2024-montpellier, KandiahPeverelle2021-accelWattch} or use machine learning (ML) models to predict energy consumption~\cite{WuGreathouse2015-amdGPUPowerML}.
Collectively, these works significantly improved the state-of-the-art in GPU energy modeling, which previously relied either on hand-rolled, non-validated power models or on validated but increasingly out-of-date power models~\cite{LengHetherington2013-gpuWattch}.

Arafa et al.'s GPU energy model~\cite{ArafaElWazir2020-nmsu} uses PTX assembly instructions to derive per-instruction energy.
Since their model is based on PTX instructions, it can adapt to different NVIDIA GPU architectures and provide fine-grained energy information.
However, because their model is based on PTX instructions, changes to compiler flags can affect what the GPU executes~\cite{GutierrezBeckmann2018-amdGem5, KhairyShen2021-accelSim}.
This constrains the model's ability to accurately predict for applications that use different compiler flags.
Moreover, the model only captures GPU compute instructions. 
On the other hand, Delestrac et al. focus on the GPU memory hierarchy's energy~\cite{DelestracMiquel2024-montpellier}.
Similar to Arafa et al., they use microbenchmarks but combine them with performance counter and sensor data to improve accuracy.
This approach provides fine-grained energy information about the memory hierarchy.
However, for workloads that are not memory-bound, the lack of information about compute instructions means that certain bottlenecks are not accounted for.
Combining these works to form a complete model is challenging.
First, since they were designed on different architecture generations, combining their results directly does not produce an accurate model.
Moreover, while combining them improves completeness, neither attributes energy to control-flow instructions, which are standard in complex applications.
Thus, these works, whether in isolation or in combination, are insufficient for energy attribution.

Conversely, Guser generates and analyzes the power behavior of individual GPU PTX instructions.
Specifically, Guser focuses on the max power per instruction~\cite{shan2024-guser}.
Unlike Arafa et al. and Delestrac et al., Guser includes both compute and memory instructions.
Although Guser's primary focus is max power per instruction, 
it also models GPU energy consumption as part of its tests.
However, like the previous works, Guser has similar limitations from its PTX-based methodologies.
Furthermore, while Guser includes both compute and memory instructions, it does not attribute energy to control-flow instructions, which are standard in complex applications.

Finally, other work uses ML to estimate the energy consumption of GPU workloads given trained estimates from other workloads~\cite{Alavani2023-gpuMLpred, WuGreathouse2015-amdGPUPowerML}.
By leveraging a large training corpus of GPU performance-counter data, it is robust across different cooling sources and vendor architectures.
However, it only provides kernel-level information about the overall behavior of a given application.
Thus, while ML models can achieve high accuracy, these models do not provide the fine-grained information needed to optimize energy consumption.
As none of these approaches provides a comprehensive picture of GPU energy consumption, we do not directly compare with them.

\subsubsection{AccelWattch}
\label{subsubsec:back-alt-accelwattch}
AccelWattch is %
the state-of-the-art in modeling GPU power consumption, frequently providing higher accuracy and robustness than the other approaches in Table~\ref{tab:relWork}.
Unlike prior approaches, AccelWattch estimates power for all instructions.
To do this, AccelWattch uses a series of microbenchmarks, in combination with quadratic programming and sensor measurements. %
Notably, because AccelWattch models power consumption, it must model energy via ``windows'' of time over which it integrates. %
Although AccelWattch purports to provide generalizable energy predictions over different GPU architectures, as shown in Section~\ref{sec:intro} and Figure~\ref{fig:accelwattch_summit}, this support is fragile, even for its modeled V100 GPU.

This fragility largely stems from AccelWattch
not being flexible enough to handle variations in architecture and operational conditions.
These differences include the GPUs having a different TDP (300W on CloudLab's V100 GPU versus 250W for AccelWattch's V100), different maximum frequencies (1417 MHz on AccelWattch's V100, 1530 MHz on CloudLab's V100), and available memory (32 GB on AccelWattch's V100, 16GB on CloudLab's V100).
To address this issue, we attempted to reproduce the temperature-aware calibration of AccelWattch on this water-cooled system.
However, AccelWattch's iterative quadratic solver-based procedure did not converge and reported zero power for data caches -- similar to prior AccelWattch reports~\cite{accelWattch-githubIssues-1, accelWattch-githubIssues-2}.
Consequently, we were also unable to create a calibrated model for this %
system.
Thus, while AccelWattch significantly improves on the state-of-the-art, it has shortcomings that make it difficult to
use in practice.

\section{Design}
\label{sec:des}

\begin{figure}[tb!]
    \centering
    \vspace{1ex}
    \includegraphics[width=0.95\linewidth]{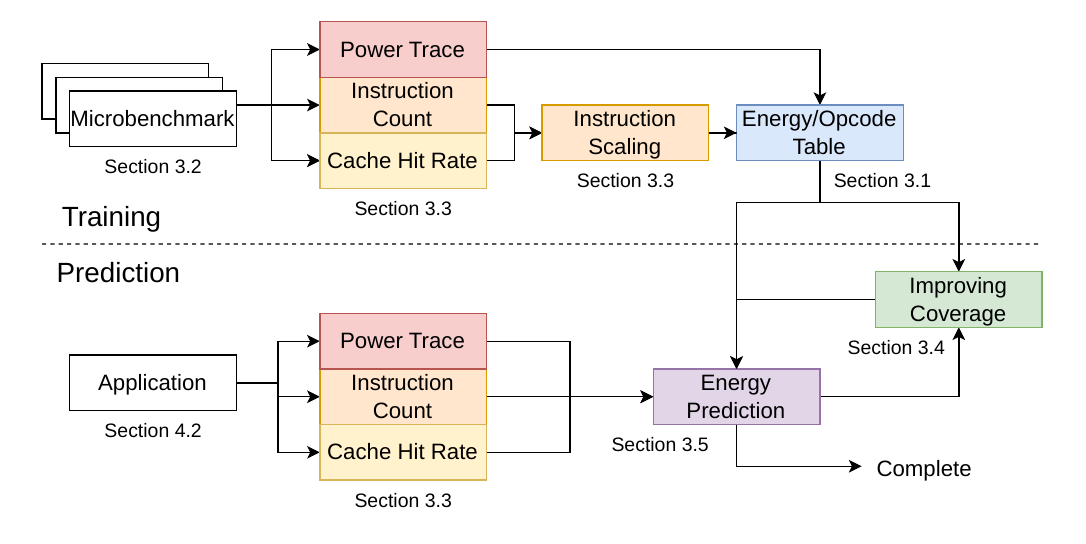}
    \vspace{-3ex}
    \caption{\DESIGN{} design overview.}
    \Description{\DESIGN{} design overview.}
    \label{fig:model-design}
    \vspace{-3ex}
\end{figure}

To overcome the inefficiencies of prior work, we exploit two key insights.
First, we observe that measuring steady-state energy consumption over long periods yields robust measurements that readily adapt to architectural and system changes, such as cooling.
Second, we solve for instruction energies as a system of equations, as ancillary instructions for one microbenchmark may be the primary instruction for another.
Figure~\ref{fig:model-design} shows \DESIGN{}'s overall design.
Overall, \DESIGN{} has two phases: a training phase (top) and a prediction phase (bottom).
The training phase uses a series of microbenchmarks (Section~\ref{subsec:des-ubmks}), profiler information, and our steady state energy measurement approach (Section~\ref{subsec:des-measure}) to isolate and create a table of per-instruction energies.
This includes hard to isolate instructions which are handled by the use of a non-negative linear solver (Section~\ref{subsec:des-nnSolve}.
In the prediction phase, \DESIGN{} uses these per-instruction energies, in combination with additional optimizations to increase instruction coverage (Section~\ref{subsec:des-coverage}), to predict the energy of GPU applications (Section~\ref{subsec:des-predict}).

\subsection{Deriving Energy Per Instruction}
\label{subsec:des-nnSolve}
A direct way to derive an energy per instruction would be to design a specific microbenchmark for each instruction; then one can measure the energy of that benchmark and amortize it across number of times the instruction was executed.
However, each microbenchmark typically requires ancillary instructions, such as load instructions, before its vector ALU or tensor operations.
Likewise, when designing tests to access various levels of the memory hierarchy, the program cannot simply load or store in isolation -- there must also be additional instruction(s) for calculating addresses and/or obtaining data values.
Since these additional instructions contribute to the overall energy, simply solving for energy based on the primary instruction can lead to over prediction.
Our key insight is that the ``extra'' instructions in one microbenchmark may be the primary instructions in another microbenchmark.
Thus, by holistically considering all microbenchmarks, we can appropriately attribute each instruction's energy contribution.

\begin{figure}[tb!]
    \centering
    \includegraphics[width=0.9\linewidth]{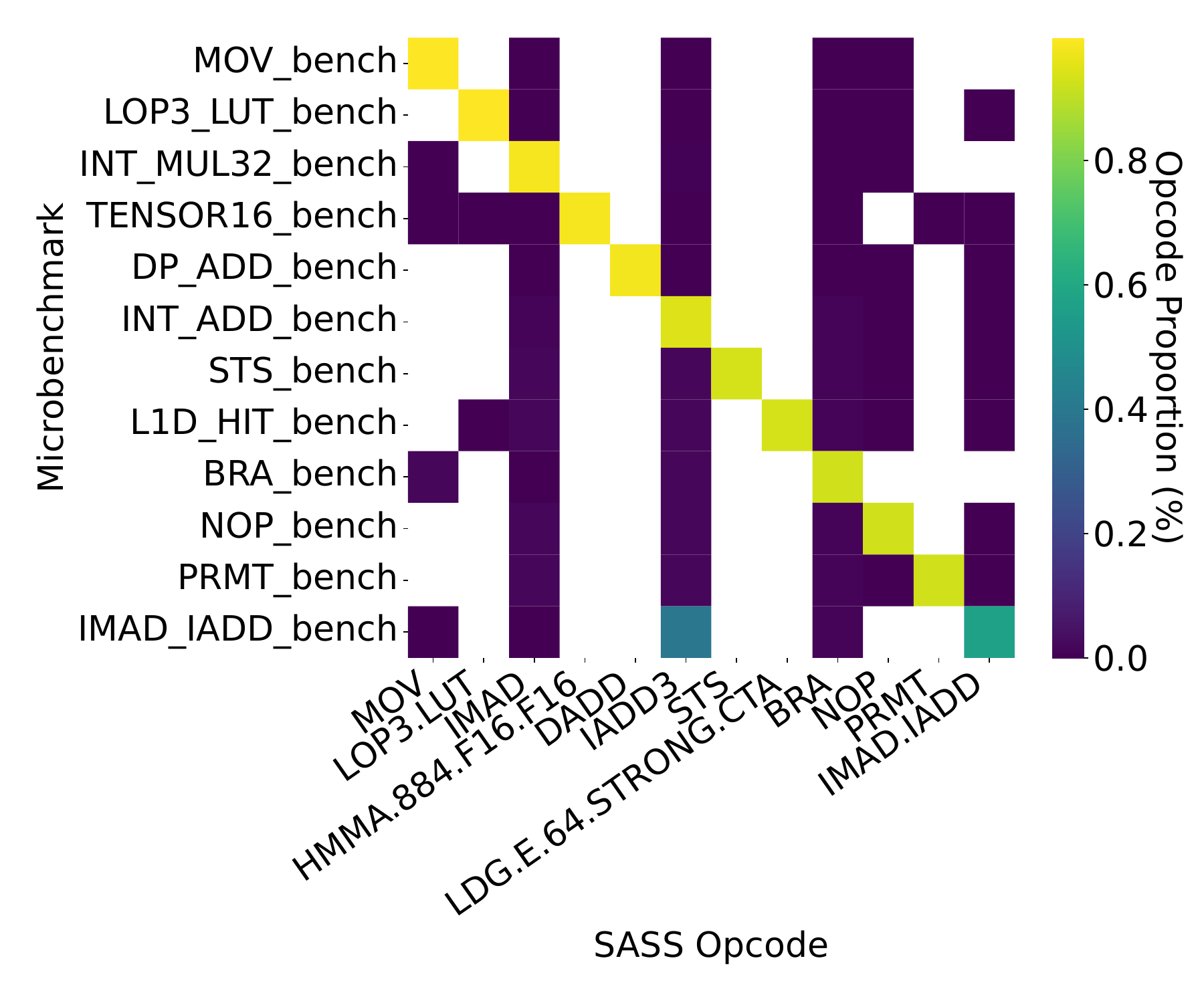}
    \vspace{-2ex}
    \caption{Subset of the full system of equations used to solve for instructions for the air-cooled V100 GPU. Each row represents one microbenchmark, and each column represents the frequency of a target instruction occurring in the benchmark (selection). The full table for the V100 GPU includes 90 microbenchmarks covering 90 instructions.}
    \Description{Subset of microbenchmark suite with their target instruction. Each row represents one microbenchmark, and each column represents the frequency with which the target instruction occurred in the benchmark (selection).}
    \label{fig:ubench_opcode_heatmap}
    \vspace{-3ex}
\end{figure}
To illustrate this, Figure~\ref{fig:ubench_opcode_heatmap} shows a subset of our microbenchmarks with the fraction of instructions in each microbenchmark.
For example, the \texttt{IMAD\_IADD\_bench} has 58\% \texttt{IMAD.IADD} instructions, 40\% \texttt{IADD3} instructions, and $<$ 1\% \texttt{MOV}, \texttt{IMAD}, and \texttt{BRA} each.
Thus \texttt{IMAD\_IADD\_bench} by itself would be a poor choice for calculating the energy consumed by \texttt{IMAD.IADD}.
However, by considering  \texttt{INT\_ADD\_bench}, we can derive an energy value for \texttt{IMAD.IADD}.
Furthermore, with the other microbenchmarks can cover the instructions such as \texttt{MOV}, \texttt{IMAD}, and \texttt{BRA}.
Hence, by solving for energies not in isolation, but as a system of equations, we can gain a more accurate estimation for the energy per instruction.
To ensure we obtain real energy per instruction values, we use a non-negative solver to enforce the non-negativity constraint.
We also maintain a square system of equations, introducing a new benchmark(s) when incorporating a new instruction(s).
We monitor the residual throughout this iterative process and observe that it remains zero, supporting our claim of a linear model for energy contributions.
We next discuss how we design our microbenchmarks and ensure stable energy measurements.

\subsection{Microbenchmark Design}
\label{subsec:des-ubmks}

\begin{lstlisting}[
    float,
    caption={Code Snapshot of Microbenchmark.},
    label={algo:microbenchmark},
    language=C++,
    basicstyle=\ttfamily\footnotesize, %
    numbers=left,                      %
    numberstyle=\tiny\color{gray},     %
    stepnumber=1,                      %
    numbersep=8pt,                     %
    xleftmargin=15pt,                  %
    framexleftmargin=15pt,             %
    frame=single,                      %
    rulecolor=\color{black!40},        %
    breaklines=true,
    keywordstyle=\color{blue}\bfseries,%
    commentstyle=\color{ForestGreen}   %
]
__global__ void shfl_kernel(unsigned* A,
                            unsigned* B,
                            unsigned long long N){
    int i = blockDim.x * blockIdx.x + threadIdx.x;
    unsigned sink = A[uid];
    #pragma unroll 100
    for(unsigned long long i=0; i<N; i++){
        sink += __shfl_sync(0xffffffff, sink, 0); 
    }
    B[uid] = sink;
}
\end{lstlisting}

Like prior work (Section~\ref{subsec:back-alt}), we design a large number of hand-tuned microbenchmarks largely using inlined assembly.
However, we focus on constructing microbenchmarks that emit target SASS instructions.
Doing so allows us to focus on what is actually being executed on the hardware and bypass any changes that may arise due to setting compiler flags.
Furthermore, we found that %
prior microbenchmarks~\cite{KandiahPeverelle2021-accelWattch} did not provide sufficient, fine-grained coverage.
For example, Listing~\ref{algo:microbenchmark} shows a microbenchmark that we added to target the \texttt{SHFL} instruction.
We also added new tests for
various data widths and levels of the GPU memory hierarchy.
Specifically, %
we create microbenchmarks that isolate the energy of memory accesses of different precisions (e.g., 8-, 16-, 32-, 64-, and 128-bits per thread), as well as each level of the memory hierarchy (L1, L2, DRAM), and utilize hit and miss profiling information, among other things, to determine a given application's energy prediction (Section~\ref{subsec:des-predict}).

To help each microbenchmark reach a stable phase, we execute the desired instruction(s) in a loop with a user-defined number of iterations.
Moreover, we unroll this loop to increase the proportion of the target instruction.
Additionally, to ensure no interference, we run each microbenchmark in isolation on the GPU and saturate the thread blocks for each microbenchmark.
This ensures the microbenchmark runs across all of the GPU's SMs and helps us distinguish the energy consumed by executing instructions from the energy attributable to shared resources.

\subsection{Ensuring Consistent and Stable Measurements}
\label{subsec:des-measure}

Unlike prior approaches (Section~\ref{subsec:back-alt}), we run workloads long enough to identify their steady-state dynamic power.
This also ensures they have a very stable power usage during their runs.
We also observe that they reach a stable temperature as the heat dissipation levels off.
Thus, our approach is agnostic towards cooling and other effects that affect the system's temperature.

\begin{figure}[tb!]
    \centering
    \vspace{-2ex}
    \includegraphics[scale=0.25]{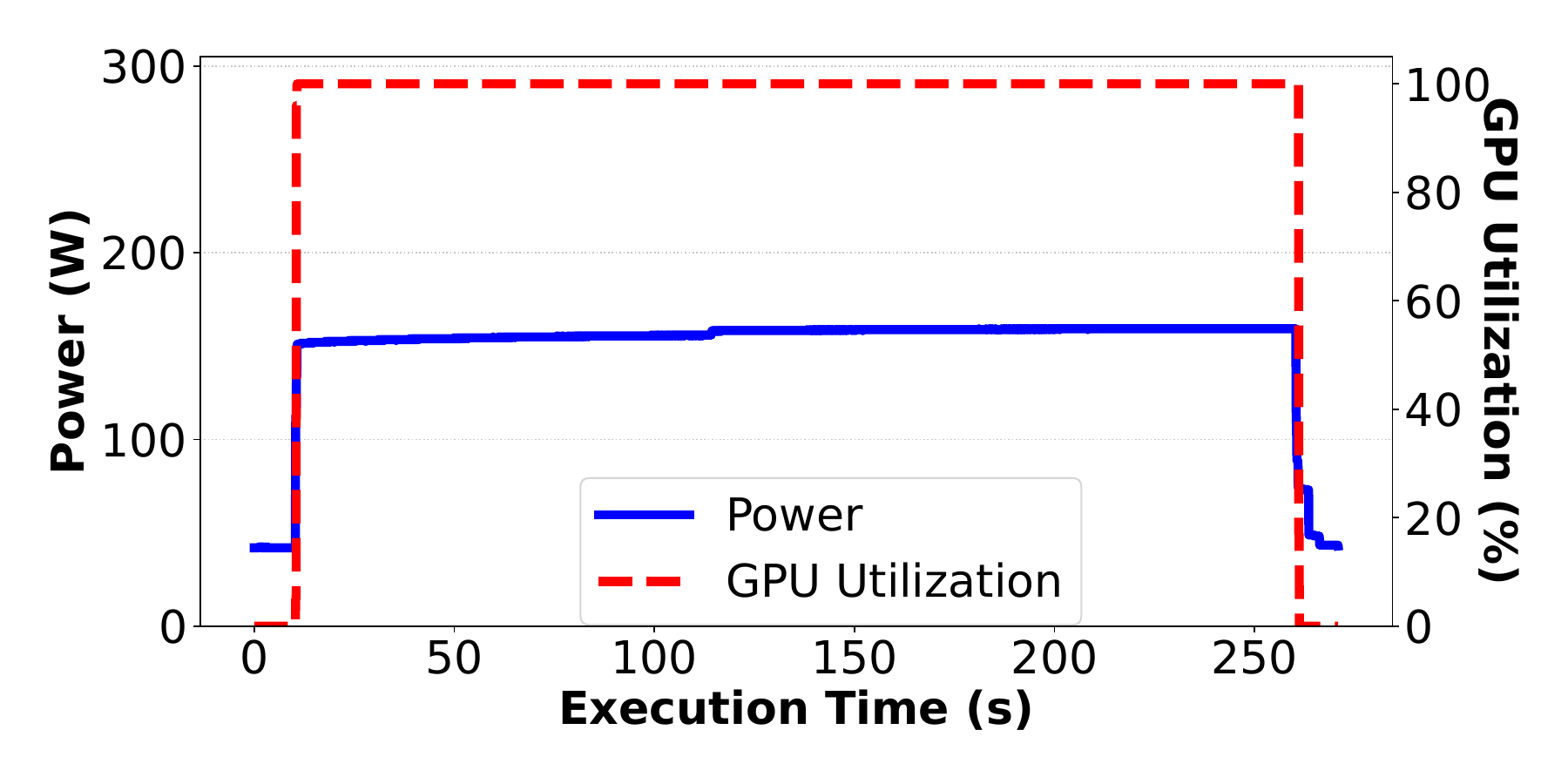}
    \caption{Power trace sampled with \texttt{NVML} from running a double precision addition microbenchmark on an air-cooled Tesla V100 GPU, including GPU utilization (red) and GPU power (blue).}
    \Description{line graph showing power samples}
    \label{fig:power_trace}
    \vspace{-4ex}
\end{figure}

Figure~\ref{fig:power_trace} shows an example of our steady state approach, which tracks the utilization (red) and power (blue) of the microbenchmark.
During our energy profiling, we also take power samples using \texttt{NVML}.
In general, we found that using an approximate integration over the power samples only differs from the values reported by the NVML energy measurement counter by less than 1\%.
Notably, after a brief startup period, Figure~\ref{fig:power_trace}'s microbenchmark remains steady at $\sim$150W.
This makes it easy to calculate a microbenchmark's total energy by summing the area under its power curve and multiplying by its execution time.
In turn, given a specific number of instances of those instructions (e.g., directly counting or from the profiler), we can accurately estimate the average energy consumption per instruction instance.
However, this energy value needs to be further subdivided into static, constant (idle), and dynamic components to correctly identify a given instruction's dynamic energy contribution.

\subsubsection{Isolating Static \& Constant Power}
\label{subsubsec:des-measure-staticIdle}
Broadly, GPU energy can be broken into constant, static, and dynamic energy (Equation~\ref{eq:totalE}).

\begin{equation}
  E_{total} = E_{const} + E_{static} + E_{dynamic}
  \label{eq:totalE}
\end{equation}

Constant energy is the amount of energy a GPU consumes in its lowest power state.
This energy is a fixed constant cost, even if no work is issued to the GPU.
Thus, constant energy can be easily calculated as the product of the constant power and the application's execution time.

The GPU's shared resources also consume static energy.
For example, for any GPU instruction, various components must be powered, such as the caches.
Even if the instruction %
does not interact with the memory system, the caches are still powered.
The amount of energy attributed to these shared resources varies with the number of active SMs and threads in a warp.
When all threads and all SMs are active, the static power is fixed to some set value, which can be multiplied by execution time to get the total energy:

\begin{equation}
  E_{total} = [(P_{const} + P_{static}) \times T_{exec}] + E_{dynamic}
  \label{eq:totalE-power}
\end{equation}

To isolate the static and constant energy from the dynamic energy, we also gather 
power samples before any application runs.
Since we control what the GPU runs, it must be idle during this period.
Thus, we can determine the GPU's constant power, convert it to energy, and subtract this energy from the energy consumed by a given microbenchmark.
Although GPU temperature fluctuates (e.g., due to prior programs), allowing the device to cool down between training experiments and taking the median across multiple runs reduces measurement variability.
Thus, we can obtain the GPU's constant power and subtract the corresponding energy. %

Furthermore, Oles, et al.~\cite{Vladyslav2024} observed that when a Volta-class V100 GPU is active but not doing work, it consumes $\sim$80W on the Summit supercomputer.
We verified that this was also true when running a \texttt{NANOSLEEP} kernel on the Volta-, Ampere-, and Hopper-class clusters we use for evaluation (Section~\ref{subsec:meth-sys}).
Based on this, we separate the static energy from the dynamic energy.
Finally, after accounting for static and constant energy, we categorize the remaining energy as dynamic energy.

\begin{figure}[tb!]
    \centering
    \includegraphics[width=0.9\linewidth]{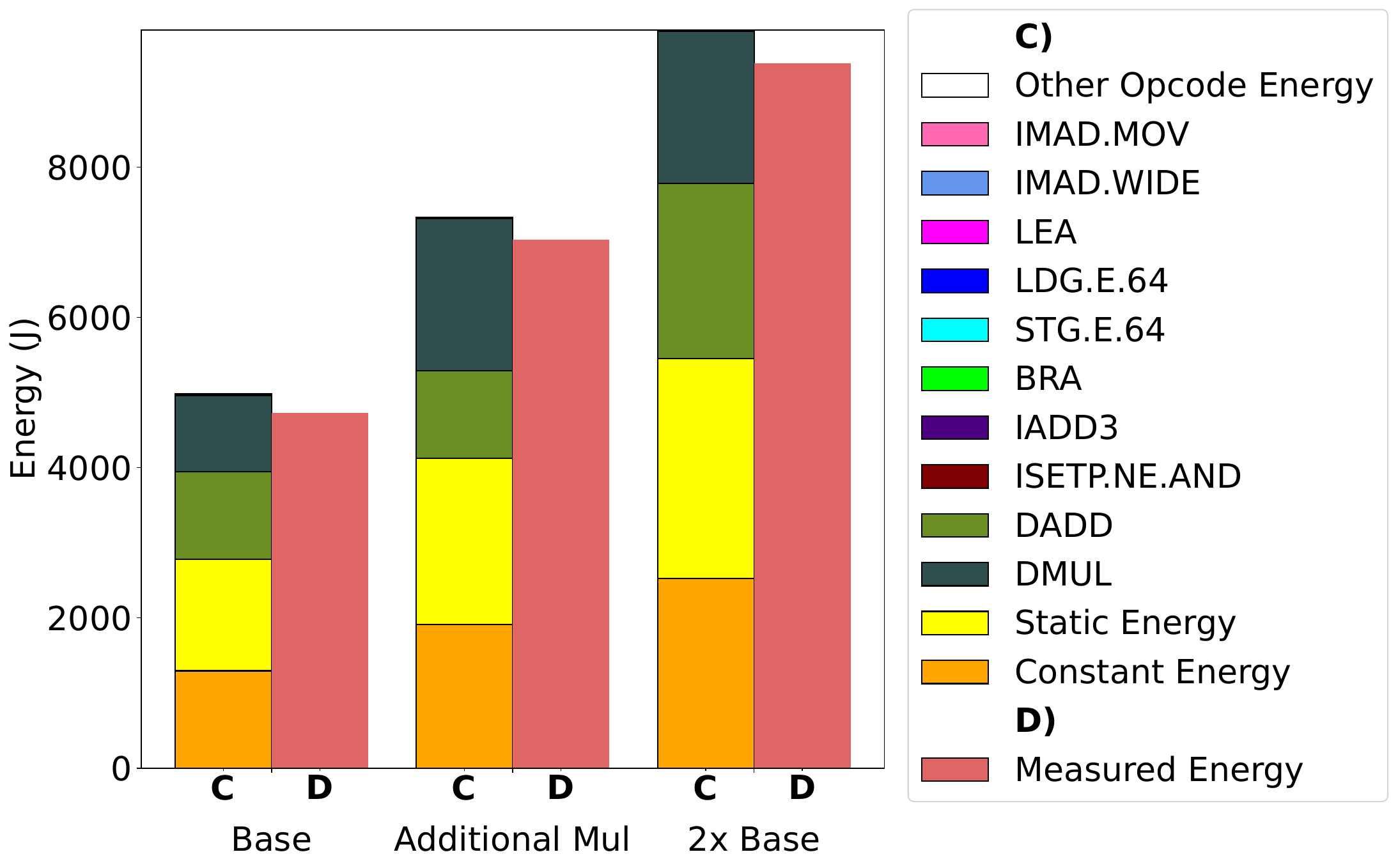}
    \vspace{-1ex}
    \caption{Simple microbenchmark that loops two different instructions. Base: 2mul and 2add; Additional Mul: 4mul and 4add; 2x Base: 4mul + 4add.}
    \Description{Simple microbenchmark that loops two different instructions. Base: 2mul and 2add; Additional Mul: 4mul and 4add; 2x Base: 4mul + 4add.}
    \label{fig:trivia_test}
    \vspace{-2ex}
\end{figure}

\noindent
\textbf{Solving for Energy:} We use dynamic energy as the right-hand side of our system of equations.
For example, in Figure~\ref{fig:trivia_test} we begin with a simple microbenchmark (labeled \textit{Base}) that executes two add instructions and two multiply instructions in a loop.
Next, we introduce two additional multiply instructions in the loop (labeled  \textit{Additional Mul}).
Lastly, we introduce two additional add instructions into the loop, resulting in twice as many instructions as the base microbenchmark (labeled 2$\times$ \textit{Base}).
As the machine executes more instructions, its runtime increases, thereby increasing the constant and static energy (Equation~\ref{eq:totalE-power}).
We observe that the remaining dynamic energy increases linearly with the instruction count.
Given this observation, we use a linear model for calculating dynamic energy (Equation~\ref{eq:dynmaicE-power}) across all \texttt{n} instructions that comprise a microbenchmark:
\begin{equation}
  E_{dynamic} = \sum^{n}_{i} count_{i} \times energy_{i}
  \label{eq:dynmaicE-power}
\end{equation}
After calculating this for all our microbenchmarks, we next develop a system of linear equations.
By solving all equations simultaneously, we can assign energy values to each instruction and account for ancillary instructions (e.g., for-loop constructs or accesses to different memory hierarchy levels).

\subsection{Improving Coverage for Low Usage Instructions}
\label{subsec:des-coverage}

Collectively, the instructions we identified energy values for (Section~\ref{subsec:des-measure}) and those our %
solver identified (Section~\ref{subsec:des-nnSolve}) enable \DESIGN{} to \textbf{directly} predict many GPU instruction's energies.
However, given the significant number of GPU instructions and modifiers to those instructions, it is difficult to measure all of them -- especially seldom-used instructions -- directly.
For these instructions, we introduce three mechanisms to \textbf{predict} energy values and increase coverage: \textit{scaling}, \textit{grouping}, and \textit{bucketing}.

\noindent
\textbf{Scaling}: Memory operations can perform the same operation but utilize different data widths or access different levels of the cache hierarchy.
Although we design unique microbenchmarks to measure the energy of many of these instruction variants directly, we also perform \textit{scaling} when microbenchmarks are unavailable. We describe how scaling is used in Section~\ref{subsec:des-predict}.

\noindent
\textbf{Grouping}: GPU ISAs frequently append several modifiers to instructions to indicate subtle differences in behavior.
For example, some memory operations will add modifiers indicating whether this address should be first for eviction, such as \texttt{STG.E.EF.64}, which is treated the same as \texttt{STG.E.64}.
Likewise, instructions that set predicate values for control flow use slightly different operations (e.g., AND, OR), such as \texttt{ISETP.GE.OR}, \texttt{ISETP.LE.AND}, \texttt{ISETP.LE.OR} are treated the same as \texttt{ISETP.GE.AND}.

This information is vital to the underlying architecture, but does not significantly affect the instruction's energy.
Thus, instead of directly writing additional tests for these instructions, we \textit{group} them by accumulating their instruction counts and assuming that all instructions in the group consume the same amount of dynamic energy per instruction, where reasonable.
We also group instruction sequences.
For example, HMMA instructions on V100 GPUs are broken into four steps, denoted with a .STEP modifier.
In such cases, we report the energy for the four instruction sequence as if it were one singular HMMA instruction.

\noindent
\textbf{Bucketing}: Prior work, like AccelWattch, categorizes GPU instructions according to microarchitectural components.
We also leverage this to predict energy for instructions that are difficult to predict directly and where other methods fail.
We first categorize the known instructions into the buckets.
For each bucket, we average all of the known energies.
Thus, for any unknown instruction $i$ that is categorized into the bucket, we assume that $i$ consumes the bucket's average energy.
For example, the \texttt{R2UR} instruction is bucketed with other typical integer ALU instructions such as \texttt{MOV} and \texttt{LOP3.LUT}.
Thus, we approximate \texttt{R2UR}'s energy as the average of the bucket's known instruction energies.

\vspace{-1ex}
\subsection{Putting It All Together: Energy Prediction \& Attribution}
\label{subsec:des-predict}

After the profiling phase, \DESIGN{} generates a constant power value, a static power value, and a table of \textbf{direct} per-instruction energies.
The \textbf{direct} per-instruction energies are categorized into groups, and then a bucket average is derived from the known energies within the bucket.

In the prediction phase, \DESIGN{} uses this data to create fine-grained energy predictions for full-sized applications.
To generate a prediction, the application is profiled to gather the application execution time, the number of instructions executed, and caching behavior.
First, the execution time is multiplied by the constant and static power values to calculate the related energy contribution.
Next, we use the profiled instruction counts to assign energies to them.
Instructions are grouped together and then checked with the per-instruction energies table, such as \texttt{ISETP.GE.OR}.

For cases where an instruction is not explicitly in the table, we use the bucket's average value for the bucket associated with that instruction, like \texttt{R2UR}.
For memory instructions, we additionally examine the cache hit rates.
For example, if the instruction is a global load like \texttt{LDG.E}, we check the profiled global load hit rate.
This allows us to incorporate the effects of instructions accessing different parts of the memory hierarchy, which are not exposed by the instruction or its modifiers.
For example, if we have an L1 Hit rate of 90\% and 100 \texttt{LDG.E} instructions, we say that 90 of them hit in the L1 and the remaining 10 missing in the L1.
Next, we compute the per-instruction energy for the instruction at the specified level of the memory hierarchy.
If the per-instruction energy at a particular level of the hierarchy is not in our table, we apply a scaling factor derived from comparing the relative energies of another instruction with known energies at the different levels of the memory hierarchy.
Similar to the compute instructions, we multiply the instruction counts by the energies and accumulate.
Thus, our result includes a prediction of total energy consumption as well as fine-grained energy attribution details.

\section{Methodology}
\label{sec:meth}

\subsection{Evaluated Systems}
\label{subsec:meth-sys}

\begin{table}[tb!]
    {\scriptsize
  \centering
  \vspace{1ex}
  \caption{Summary of studied GPUs and HPC clusters.}
  \vspace{-2ex}
  \begin{center}
  \begin{tabular}{|l|c|c|c|c|}
    \hline
    \textbf{Cluster} & \textbf{GPU} & \textbf{Cooling} \\ \hline
    \textbf{CloudLab~\cite{DuplyakinRicci2019-cloudlab}} & V100 & air \\ \hline
    \textbf{Summit}~\cite{ornl} & V100 & water \\ \hline
    \textbf{Lonestar6}~\cite{tacc} & A100 & air \\ \hline
    \textbf{Lonestar6}~\cite{tacc} & H100 & air \\ \hline
    \end{tabular}
  \end{center}
  \vspace{-1ex}
  \label{tab:hpc-clusters}
  \vspace{-3ex}
}

\end{table}

\begin{table}[tb!]
    {\scriptsize
  \centering
  \caption{Summary of studied workloads.}
  \vspace{-2ex}
  \begin{center}
  \begin{tabular}{|l|c|}
    \hline
    \textbf{Application}                                                           & \textbf{Input}  \\ \hline
    Backprop\_k1~\cite{CheBoyer2009-rodinia, CheSheaffer2010-rodinia}              & 64K \\ \hline
    Backprop\_k2~\cite{CheBoyer2009-rodinia, CheSheaffer2010-rodinia}              & 64K \\ \hline
    Hotspot~\cite{CheBoyer2009-rodinia, CheSheaffer2010-rodinia}                   & 1024 2 20 temp\_1024 power\_1024 \\ \hline
    KMeans~\cite{CheBoyer2009-rodinia, CheSheaffer2010-rodinia}                    & 819200 \\ \hline
    SRAD\_v1~\cite{CheBoyer2009-rodinia, CheSheaffer2010-rodinia}                  & 100, 0.5, 502, 458 \\ \hline
    GEMM\_c1 (Double, Float, Half)~\cite{Narang2016, Narang2017-deepBench}                              & 1760x128x1760 \\ \hline
    GEMM\_c2 (Double, Float, Half)~\cite{Narang2016, Narang2017-deepBench}                                & 3072x128x1024 \\ \hline
    RNN Training (Double, Float)~\cite{Narang2016, Narang2017-deepBench}                                   & Vanilla, 1760 hidden layers, \\
 & 16 batch size, 50 steps \\ \hline
    RNN Inference (Double, Float, Half)~\cite{Narang2016, Narang2017-deepBench}                                    & Vanilla, 1760 hidden layers, \\
 & 16 batch size, 50 steps \\
 \hline
 PageRank~\cite{PageRankSpMV} & $659033\times659033$\\ \hline %
 QMCPack~\cite{Godoy2025Softwarestewardship-QMCPACK, Kim2018, Kent2020}           & NiO S64 (256 atoms, 3072 edges) \\ \hline
  \end{tabular}
  \end{center}
  \vspace{-1ex}
  \label{tab:bmks}
  \vspace{-3ex}
}

\end{table}

To ensure our results were representative across a variety of GPU clusters and systems, %
we ran our experiments across clusters with different sizes, generations, and cooling approaches.
Table~\ref{tab:hpc-clusters} summarizes the four HPC clusters %
we studied.
We use TACC's modern Lonestar6 cluster, which includes 252 air-cooled NVIDIA A100 and eight air-cooled H100 GPUs~\cite{tacc}.
To demonstrate \DESIGN{}'s generalizability, we studied ORNL's Summit supercomputer, a large, water-cooled cluster with 27648 NVIDIA V100 GPUs.
CloudLab is a smaller, air-cooled cluster with 12 NVIDIA V100 GPUs~\cite{DuplyakinRicci2019-cloudlab}.
For our V100 study, we also provide energy estimates from AccelWattch, as it has validated power models for these GPUs.

\subsection{Applications}
\label{subsec:meth-apps}

We evaluate \DESIGN{} across 90 microbenchmarks and 16 traditional GPGPU~\cite{CheBoyer2009-rodinia, CheSheaffer2010-rodinia}, graph analytics~\cite{PageRankSpMV}, ML~\cite{Narang2016, Narang2017-deepBench}, and HPC~\cite{Kim2018,Kent2020,Godoy2025Softwarestewardship-QMCPACK}, workloads with diverse behavior.
This mix is representative of how modern systems are used and stresses a variety of GPU components, including ALUs, specialized memories, TensorCores, and various levels of the memory hierarchy.
Table~\ref{tab:bmks} summarizes these workloads.

\noindent
\textbf{Microbenchmarks}:
We use a suite of 90 microbenchmarks, 51 of which we wrote to augment AccelWattch's existing microbenchmarks.
Each microbenchmark uses inlined assembly to isolate the behavior of specific GPU components, including shared memory, cache levels, main memory, TensorCores, and various ALU and control flow operations.
The microbenchmark saturates the GPU's SMs and, within a thread block, saturates all SIMT lanes per SM.
We do not show results for the microbenchmarks in Section~\ref{sec:eval} as they are inputs for \DESIGN{}'s predictions (Section~\ref{subsec:des-predict}).

\begin{figure*}[tb!]
    \centering
    \begin{minipage}[c]{0.68\textwidth}
        \centering
        \includegraphics[width=\linewidth]{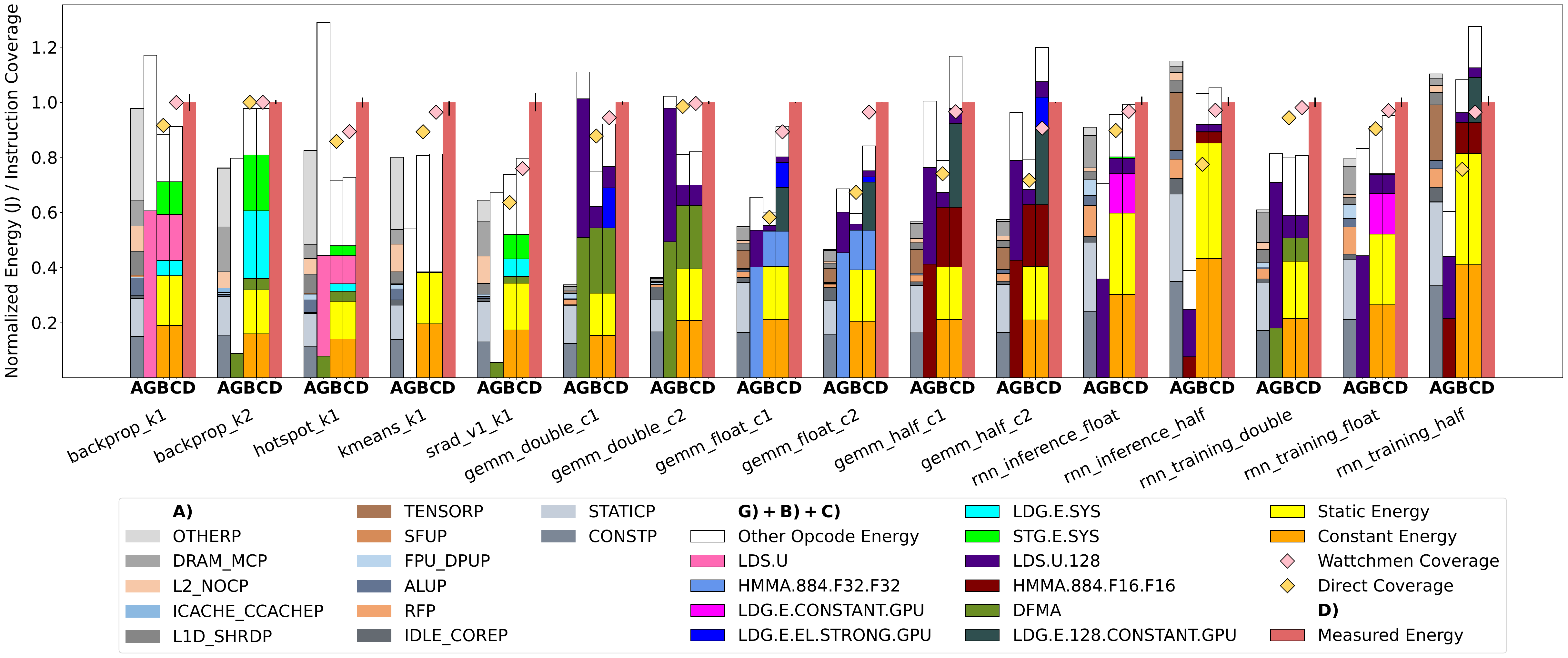}
        \captionof{figure}{Normalized energy predictions on air-cooled GPUs in CloudLab. Columns per workload: A) AccelWattch; G) Guser; B) \DESIGN{}-Direct; C) \DESIGN{}-Pred; D)~measured energy.}
        \Description{Normalized energy predictions on air-cooled GPUs in CloudLab. Columns per workload: A) AccelWattch; G) Guser; B) \DESIGN{}-Direct; C) \DESIGN{}-Pred; D)~measured energy.}
        \label{fig:cloudlab_prediction}
    \end{minipage}%
    \hfill%
    \begin{minipage}[c]{0.28\textwidth}
        \centering
        \captionof{table}{Air-cooled V100 energy estimation MAPE (\%).}
        \Description{Air-cooled V100 energy estimation MAPE (\%).}
        \label{tab:cloudlab_mape_summary}
        \vspace{1ex} %
        \begin{tabular}{lc}
            \hline
            \textbf{Model} & \textbf{MAPE (\%)} \\ \hline
            AccelWattch & 32 \\
            Guser & 25 \\
            \DESIGN{}-Direct & 19 \\
            \DESIGN{}-Predict & 14 \\ \hline
        \end{tabular}
    \end{minipage}
\end{figure*}

\noindent
\textbf{Benchmarks}:
To compare (Section~\ref{subsec:meth-configs}) against AccelWattch's average kernel power values, we alter Rodinia's GPGPU benchmarks to repeat a particular target kernel multiple times.
This enables the particular kernel to dominate the energy consumption and ensure a fair comparison.
For the graph analytics, HPC, and ML benchmarks, we keep the benchmarks as they were originally written.
For each GEMM, we study both half and full precision; for the RNNs, we study half, float, and double precisions for training and half and float precisions for inference.
For PageRank, we use the common Sparse-matrix-Vector (SPMV) algorithm~\cite{CheBeckmann2013-pannotia, PageRankSpMV}.
In particular, we use \texttt{pre2} as the input graph to demonstrate \DESIGN{} for a workload that is memory-bandwidth-bound~\cite{pre2-ATandT}.
Finally, QMCPACK is an open-source software for ab initio electronic structure calculations using real-space quantum Monte Carlo methods for solids, molecules, atoms, and model systems~\cite{Kim2018, Kent2020}.
This application is important for the DOE’s materials science, nanoscience, and quantum information science by enabling highly accurate simulations of materials and electronic structure.
Thus, QMCPACK is included in many of DOE’s application readiness efforts, such as CORAL2~\cite{coral2_benchmark}, Exascale Computing Project (ECP)~\cite{ecp_QMCPACK}, and OLCF6~\cite{olcf6_benchmark}.
This benchmark is sensitive to floating point computation, memory bandwidth, and memory latency.

For all experiments, we run the workloads on a single GPU to perform a direct comparison with AccelWattch.
However, our approach is also useful more broadly for clusters, because optimizing energy within a GPU also improves the overall energy consumption~\cite{TschandRajan2025-mlperfPower}, as node-level performance optimization is the primary step towards efficient and scalable applications.

\noindent
\textbf{Compilation}:
For all experiments across V100 GPUs, we use CUDA 11.0.
When working with A100 and H100 GPUs, we use CUDA 12.0. %
To determine the SASS instruction breakdown for each workload, we use \texttt{NSight Compute}'s SASS opcode count.
For these breakdowns, we 
retain all modifier information as these modifiers impact how the opcode is executed on the GPU.
For example, the datatype conversion instruction \texttt{F2F} (Floating Point To Floating Point Conversion) can have different modifiers depending on the conversions performed, such as \texttt{F2F.F16.F32} and \texttt{F2F.F32.F16}.

\subsection{Configurations}
\label{subsec:meth-configs}

We evaluate the following configurations (labels in parentheses are used in the figures in Section~\ref{sec:eval}):

\noindent
\textbf{AccelWattch (A)}: The state-of-the-art GPU power model (Section~\ref{subsec:back-alt}).  We used AccelWattch's publicly available V100 %
model 
to predict energy for the workloads in Section~\ref{sec:eval}.
Since AccelWattch predicts power, we converted its predictions to energy by multiplying the reported average power of a given kernel by the observed execution time.

\noindent
\textbf{Guser (G)}~\cite{shan2024-guser}:
Since the Guser model is not publicly available, we used our microbenchmark suite and apply Guser's methodology.
In particular, we take the maximum power and multiply it by the execution time, rather than integrating a steady-state power trace.
We also amortize the total energy.

\noindent
\textbf{\DESIGN{}-Direct (B)}: Our \DESIGN{} model (Section~\ref{sec:des}) which predicts energy from directly identified instructions.

\noindent
\textbf{\DESIGN{}-Pred (C)}: The extended version of \DESIGN{} with increased instruction coverage (Section~\ref{subsec:des-coverage}).

\noindent
\textbf{Real GPU (D)}: The actual measured energy for a given GPU kernel, as reported by \texttt{NVML}.

\section{Evaluation}
\label{sec:eval}

\begin{figure}[tb!]
    \centering
    \includegraphics[width=0.9\linewidth]{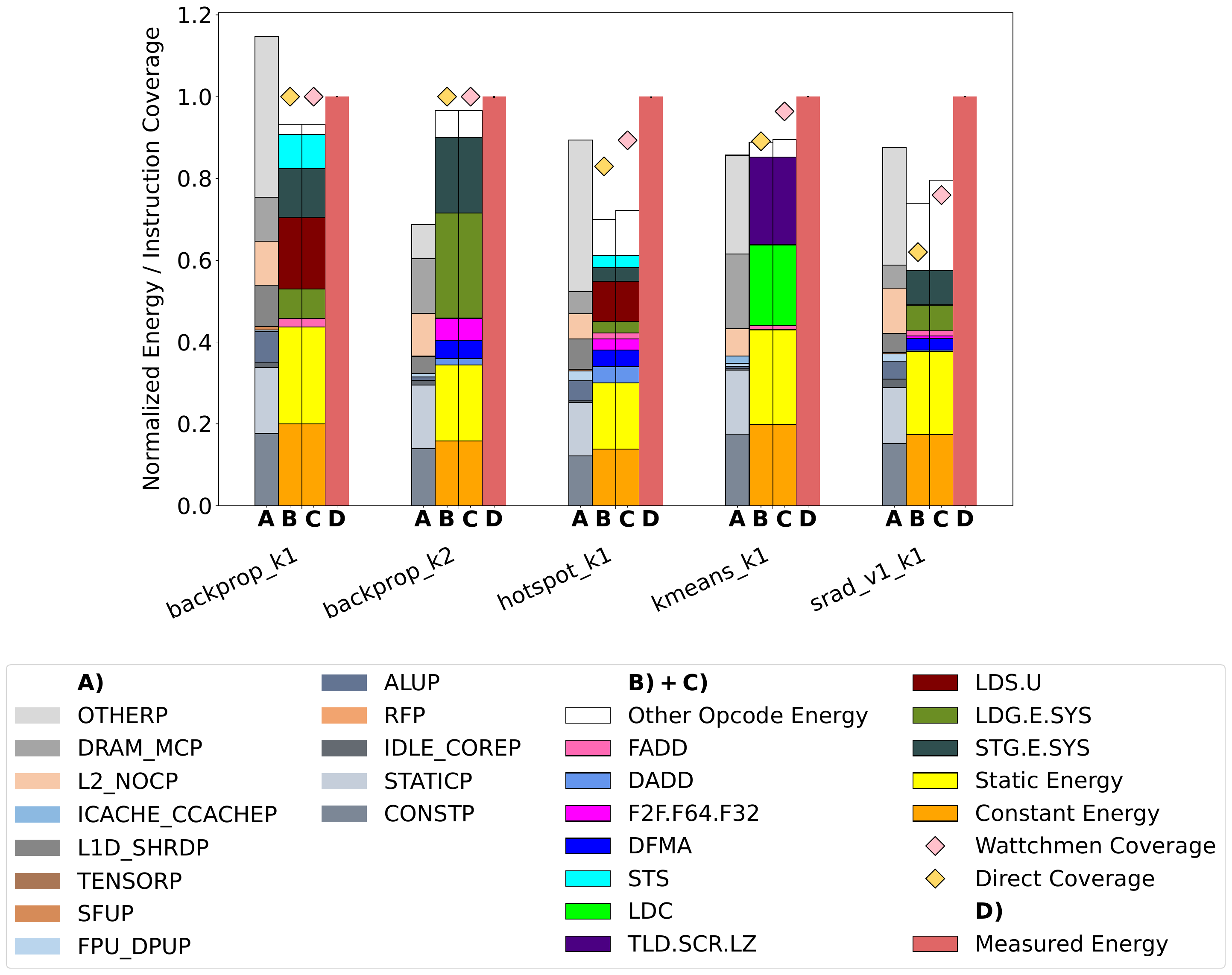}
    \caption{Applying \DESIGN{} to water-cooled V100 GPUs. Columns per workload: A) AccelWattch ; B) \DESIGN{}-Direct ; C) \DESIGN{}-Pred ; D)~measured energy.}
    \Description{Applying \DESIGN{} to water-cooled V100 GPUs. Columns per workload: A) AccelWattch ; B) \DESIGN{}-Direct ; C) \DESIGN{}-Pred ; D)~measured energy.}    
    \label{fig:summit_rodinia_prediction}
    \vspace{2ex} 
    
    \captionof{table}{Water-cooled V100 energy estimation MAPE (\%).}
    \label{tab:summit_mape_summary}
    \vspace{1ex}
    \begin{tabular}{lc}
        \hline
        \textbf{Model} & \textbf{MAPE (\%)} \\ \hline
        AccelWattch & 17 \\
        \DESIGN{}-Direct & 15 \\
        \DESIGN{}-Predict & 14 \\ \hline
    \end{tabular}
    \vspace{-4ex}
\end{figure}

Overall, our experiments show that \DESIGN{} reduces prediction error compared to AccelWattch on V100 GPUs: from 32\% to 19\% with \DESIGN{}-Direct and 14\% with \DESIGN{}-Pred.
\DESIGN{} also provides better MAPE than Guser (25\%),  further demonstrating the value in \DESIGN{}'s approach.
For water-cooled V100 GPUs (15\%), air-cooled A100 GPUs (11\%), and air-cooled H100 GPUs (12\%), \DESIGN{}'s MAPE remains similar (Section~\ref{subsec:eval-portability}) -- demonstrating its generalizability.
Finally, our case studies show how applying \DESIGN{} to HPC systems improves both Backprop's (16\%) and QMCPACK's (35\%) energy efficiency (Section~\ref{subsec:eval-casestudy}).

\subsection{Comparison With State-of-the-Art}
\label{subsec:eval-accuracy}

Since AccelWattch was validated on air-cooled V100 GPUs, we compare \DESIGN{} on the same GPUs.
Figure~\ref{fig:cloudlab_prediction} shows that AccelWattch provides a MAPE of 32\% (higher than AccelWattch's reported V100 MAPE, as discussed in Section~\ref{subsubsec:back-alt-accelwattch}).
Both \DESIGN{}-Direct (19\% MAPE) and \DESIGN{}-Pred (14\% MAPE) improve on AccelWattch.
In particular, AccelWattch struggles when predicting the energy of GEMM workloads.
This is in part due to AccelWattch's low predictions for the respective matrix and memory operations.
Conversely, \DESIGN{}-Direct properly attributes energy use of the matrix operations, and \DESIGN{}-Pred fills the gap for memory operations that were not explicitly measured. %
However, because the memory instruction's predictions rely on scaling factors rather than explicit measurements, they can overpredict, as happens with half-precision GEMMs.
The half-precision RNNs show another overprediction source. %
Compared with other workloads, RNNs have a much higher proportion of static and constant energy ($\approx$80\%), whereas in the other workloads, static and constant energy account for only $\approx$40\% of the total.
This happens because RNNs do not utilize GPUs efficiently~\cite {PatiAga2022-demystifying, ShoeybiPatwary2019-megatronlm, ZadehPoulos2019-dlTime}, causing the GPU to run at lower power.
When this happens, \DESIGN{} is more prone to error since the static and constant energy consume larger fractions of total energy.
Nonetheless, compared with the state-of-the-art AccelWattch, \DESIGN{} achieves higher accuracy in attributing energy consumption.

\noindent
\textbf{Guser Comparison}:
Figure~\ref{fig:cloudlab_prediction} shows that Guser  provides 25\% MAPE -- worse than \DESIGN{} (14\%) but outperforming AccelWattch (32\%).
Broadly, Guser works well for tests that maximize the GPU's power consumption (one of Guser's original goals). %
However, there are several factors that make it ill-suited for generating a fine-grained energy breakdown.
First, since Guser focused on max power, it ignores constant and static energy consumption.
Consequently, those costs are amortized into instruction-level values -- and causing overpredictions.
Furthermore, Guser does not factor in the energy consumption from ancillary instructions, as the impact remains consistent across its microbenchmarks.
Leaving out their impact can skew the overall energy.
Lastly, Guser is primarily focused on compute instructions.
Although Guser incorporates memory instructions in its construction, the focus on compute instructions reduces prediction accuracy for memory-intensive applications.
Overall, this demonstrates how \DESIGN{} also significantly outperforms Guser. %

\subsection{Generalization to Cooling and Later Architectures}
\label{subsec:eval-portability}

\begin{figure*}[tb!]
    \centering
    \begin{minipage}[c]{0.68\textwidth}
        \centering
        \includegraphics[width=\linewidth]{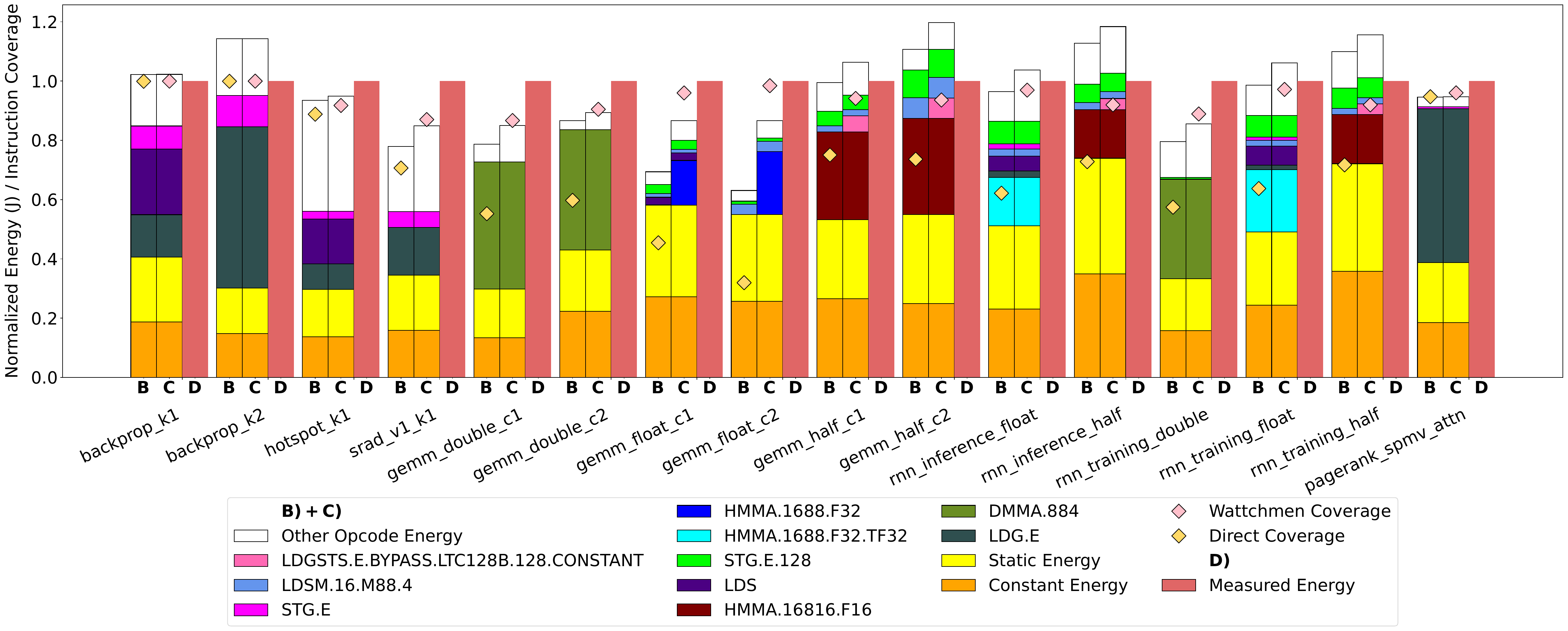}
        \captionof{figure}{Normalized energy breakdown and instruction coverage across an A100 GPU}
        \Description{Normalized energy breakdown and instruction coverage across an A100 GPU}
        \label{fig:a100_prediction}
    \end{minipage}%
    \hfill%
    \begin{minipage}[c]{0.28\textwidth}
        \centering
        \captionof{table}{Air-cooled A100 energy estimation MAPE (\%).}
        \Description{Air-cooled A100 energy estimation MAPE (\%).}
        \label{tab:a100_mape_summary}
        \vspace{1ex} %
        \begin{tabular}{lc}
            \hline
            \textbf{Model} & \textbf{MAPE (\%)} \\ \hline
            \DESIGN{}-Direct & 13 \\
            \DESIGN{}-Predict & 11 \\ \hline
        \end{tabular}
    \end{minipage}
\end{figure*}

\subsubsection{Water-cooled V100}
\label{subsubsec:eval-summit}

Next, we evaluate \DESIGN{} on the Summit supercomputer's water-cooled V100 GPUs.
Our experiments found that across the five Rodinia benchmarks, the water-cooled V100 GPUs had an average 12\% lower energy consumption than air-cooled V100 GPUs.
Interestingly, Figure~\ref{fig:summit_rodinia_prediction} shows that AccelWattch has a 17\% MAPE, lower than AccelWattch's error in the air-cooled V100 system.
Upon closer inspection, we found that this was because AccelWattch, unaware of the cooling mechanism, predicted the same energy for both systems.
However, the ground-truth energies were lower on the water-cooled Summit GPUs, which serendipitously reduces their MAPE.
Despite this, \DESIGN{}-Direct and \DESIGN{}-Pred still provides small improvements (15\%, 14\% MAPE).
This highlights one of \DESIGN{}’s key benefits over alternatives such as AccelWattch, which struggles to account for factors such as different cooling environments.
Thus, even with different cooling, \DESIGN{} again improves on AccelWattch.

\begin{figure*}[tb!]
    \centering
    \begin{minipage}[c]{0.68\textwidth}
        \centering
        \includegraphics[width=\linewidth]{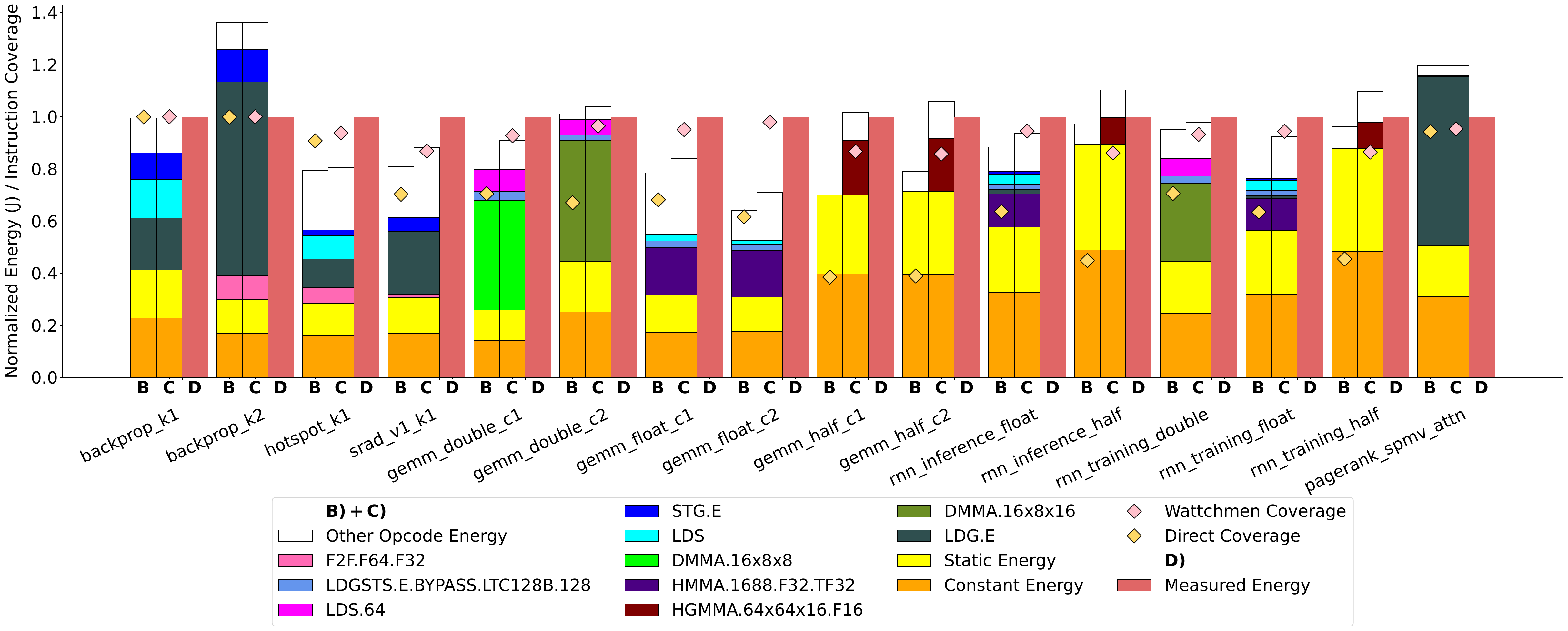}
        \captionof{figure}{Normalized energy breakdown and instruction coverage across an H100 GPU}
        \Description{Normalized energy breakdown and instruction coverage across an H100 GPU}
        \label{fig:h100_prediction}
    \end{minipage}%
    \hfill%
    \begin{minipage}[c]{0.28\textwidth}
        \centering
        \captionof{table}{Air-cooled H100 energy estimation MAPE (\%).}
        \Description{Air-cooled H100 energy estimation MAPE (\%).}
        \label{tab:h100_mape_summary}
        \vspace{1ex} 
        \begin{tabular}{lc}
            \hline
            \textbf{Model} & \textbf{MAPE (\%)} \\ \hline
            \DESIGN{}-Direct & 16 \\
            \DESIGN{}-Predict & 12 \\ \hline
        \end{tabular}
    \end{minipage}
\end{figure*}

\subsubsection{A100}
For our A100 experiments, we follow our methodology and retrain the model.
As CUDA 12.0 deprecated textures, we omit results for \texttt{kmeans\_k1}.
To compensate we introduce results for PageRank.
\DESIGN{} works well for PageRank, demonstrating its ability to accurately predict energy for not only compute-bound jobs like GEMMs, but also irregular, memory-bound workloads.
Overall, Figure~\ref{fig:a100_prediction} shows that \DESIGN{}-Direct has a 13\% MAPE across all benchmarks on A100 GPUs.
Part of \DESIGN{}-Direct's error is due to low coverage for instructions not explicitly covered in \DESIGN{}'s microbenchmark suite.
On average, \DESIGN{}-Direct attributes energy for 70\% of the instructions in the workloads.
However, \DESIGN{}-Pred overcomes this by approximating unknown instructions, increasing coverage to 93\% and reducing MAPE to 11\%.
Thus, \DESIGN{} provides high fidelity predictions without requiring microbenchmarks for every SASS instruction.

\subsubsection{H100}
\label{subsubsec:eval-h100}
NVIDIA's H100 GPUs introduce a variety of new instructions, including warp-group matrix multiply instructions (\texttt{HGMMA}).
Consequently, Figure~\ref{fig:h100_prediction} \DESIGN{}-Direct's MAPE is 16\% with a 66\% coverage.
However, bucketing again addresses this issue, raising instruction coverage to 92\%.
Accordingly, \DESIGN{}-Pred's MAPE is 12\%.
For example, the half-precision GEMMs use the new warp-group instruction \texttt{HGMMA.64x64x16.F16}.
Since \DESIGN{}-Direct does not have a microbenchmark for this, it has very low energy attribution.
Nonetheless, bucketing's first-order approximation in \DESIGN{}-Pred greatly increases coverage and improves overall energy predictions. 
Overall, across all evaluated cooling methods and architectures, \DESIGN{}-Pred's MAPE is consistently below 15\%, highlighting \DESIGN{}'s consistency across cooling methods and architectures.

\subsection{Case Studies}
\label{subsec:eval-casestudy}

\subsubsection{Case Study: Backprop}
\label{subsubsec:eval-backprop}
\begin{figure}
    \centering
    \vspace{-2ex}
    \includegraphics[scale=0.2]{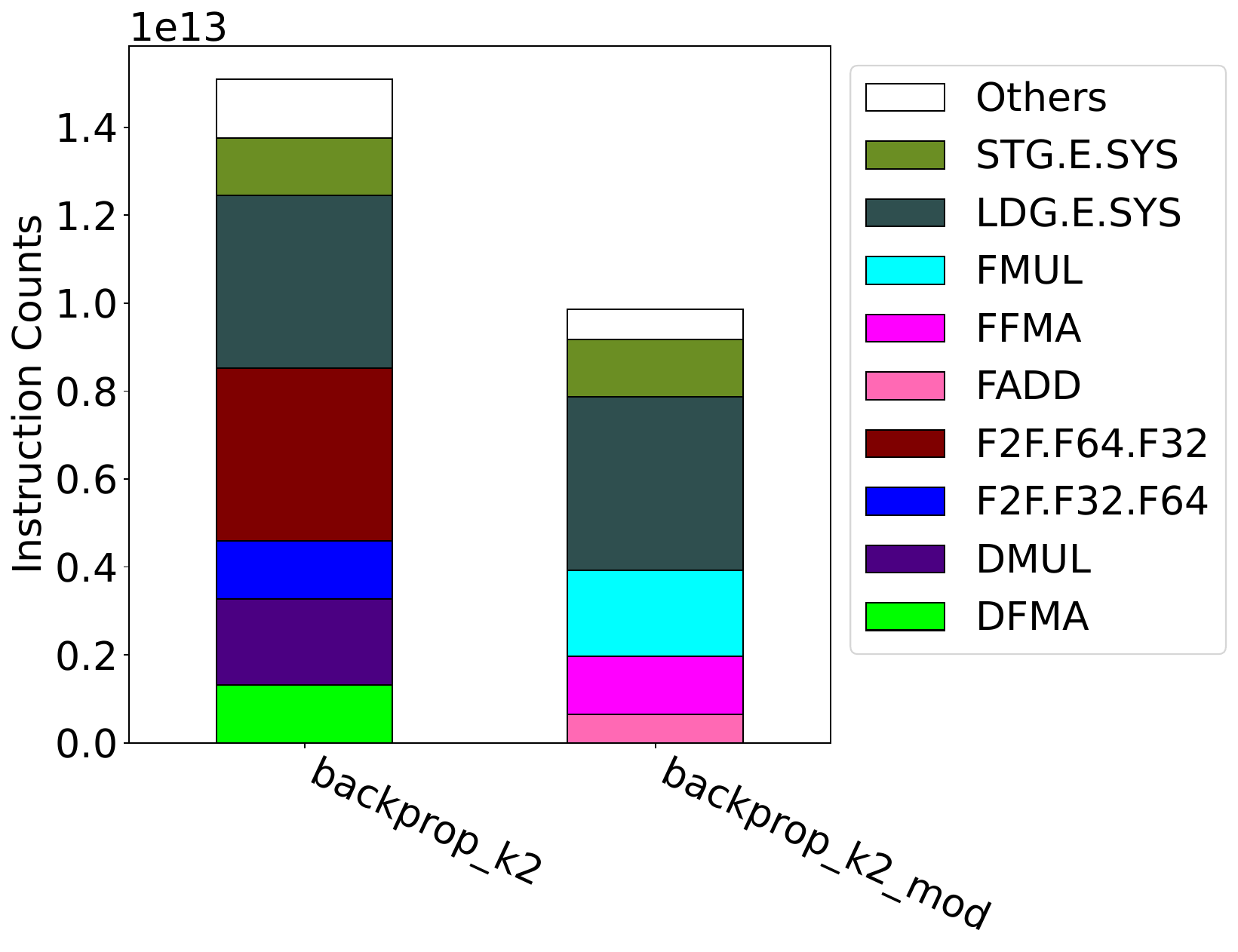}
    \vspace{-1ex}
    \caption{Opcode counts for \texttt{backprop\_k2} kernel before and after modification on an air-cooled V100 GPU.}
    \Description{Stacked graph of instruction counts}
    \label{fig:backprop_k2_observation}
    \vspace{-2ex}
\end{figure}

\begin{figure}
    \centering
    \includegraphics[scale=0.2]{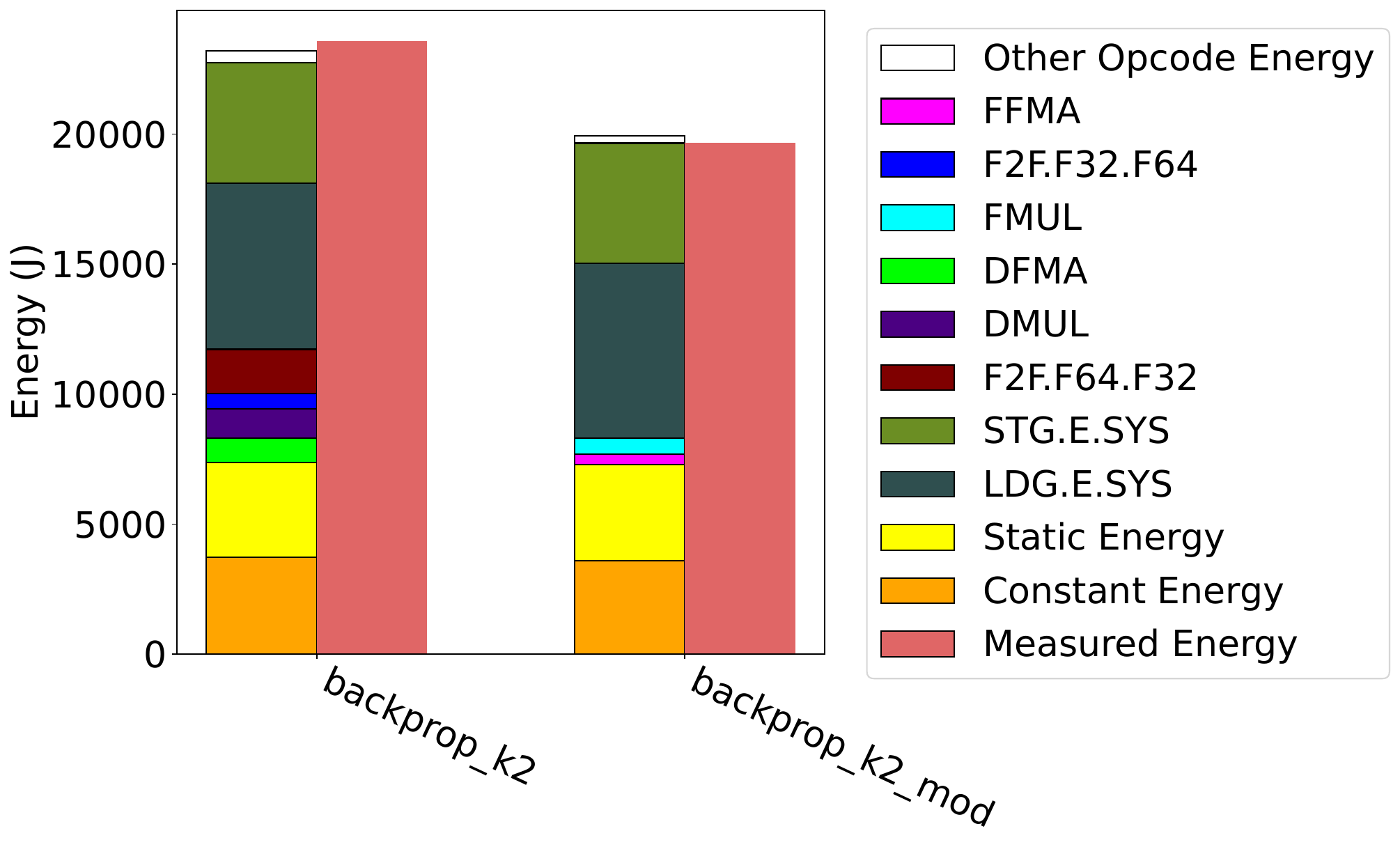}
    \vspace{-1ex}
    \caption{Energy prediction for \texttt{backprop\_k2} before and after the modification on an air-cooled V100 GPU.}
    \Description{Stacked graph of different types of predictions compared to the ground truth for different benchmarks.}
    \label{fig:backprop_k2_mod}
    \vspace{-3ex}
\end{figure}

To demonstrate how \DESIGN{} can be used to improve energy, we examine \texttt{backprop\_k2}.
In this kernel, we found that after the memory instructions, the next highest energy consuming instruction is \texttt{F2F.F64.F32}.
Figure~\ref{fig:backprop_k2_observation} shows that this instruction makes up $\approx$25\% of all \texttt{backprop\_k2} instructions.
Upon deeper investigation, 
we found that this was due to an initialization error: two $\#$\texttt{define} values used the system default of double precision, despite the operations being single precision.
Changing the defined values reduced \texttt{backprop\_k2}'s energy by 16\% (Figure~\ref{fig:backprop_k2_mod}) and improved its performance by 1\%.
Thus, this demonstrates how \DESIGN{}'s fine-grained energy profile can be used to investigate and optimize workloads on modern HPC systems.

\subsubsection{Case Study: QMCPACK}
\label{subsubsec:eval-qmcpack}

\begin{figure}[tb!]
    \centering
    \vspace{-4ex}
        \begin{subfigure}{0.9\columnwidth}
        \includegraphics[scale=0.23]{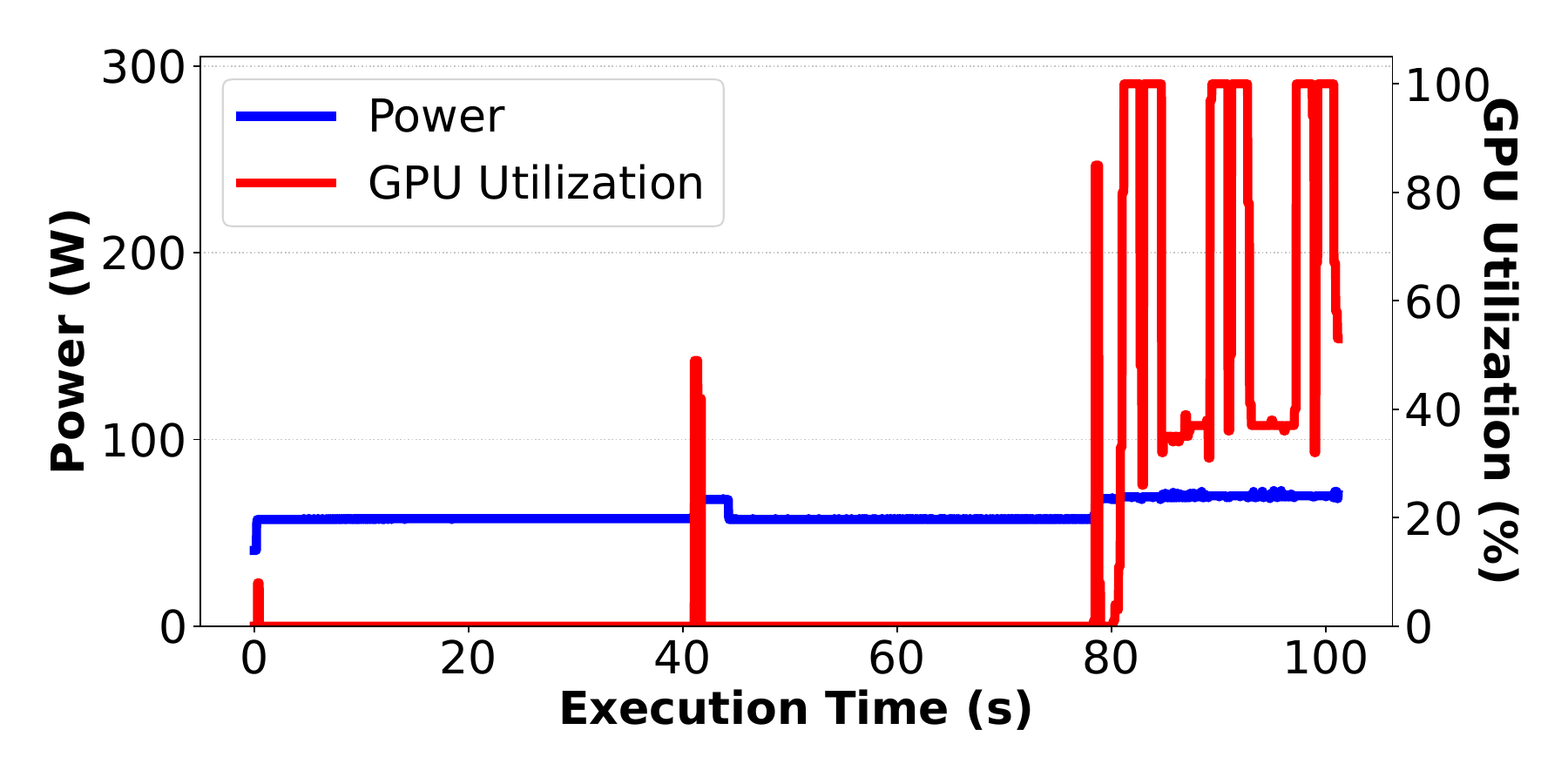} 
        \caption{Power trace from running mixed-precision QMCPACK on an air-cooled V100 GPU.} %
        \label{fig:qmcpack-mix-original}
        \end{subfigure}
        \begin{subfigure}{0.9\columnwidth}
        \includegraphics[scale=0.23]{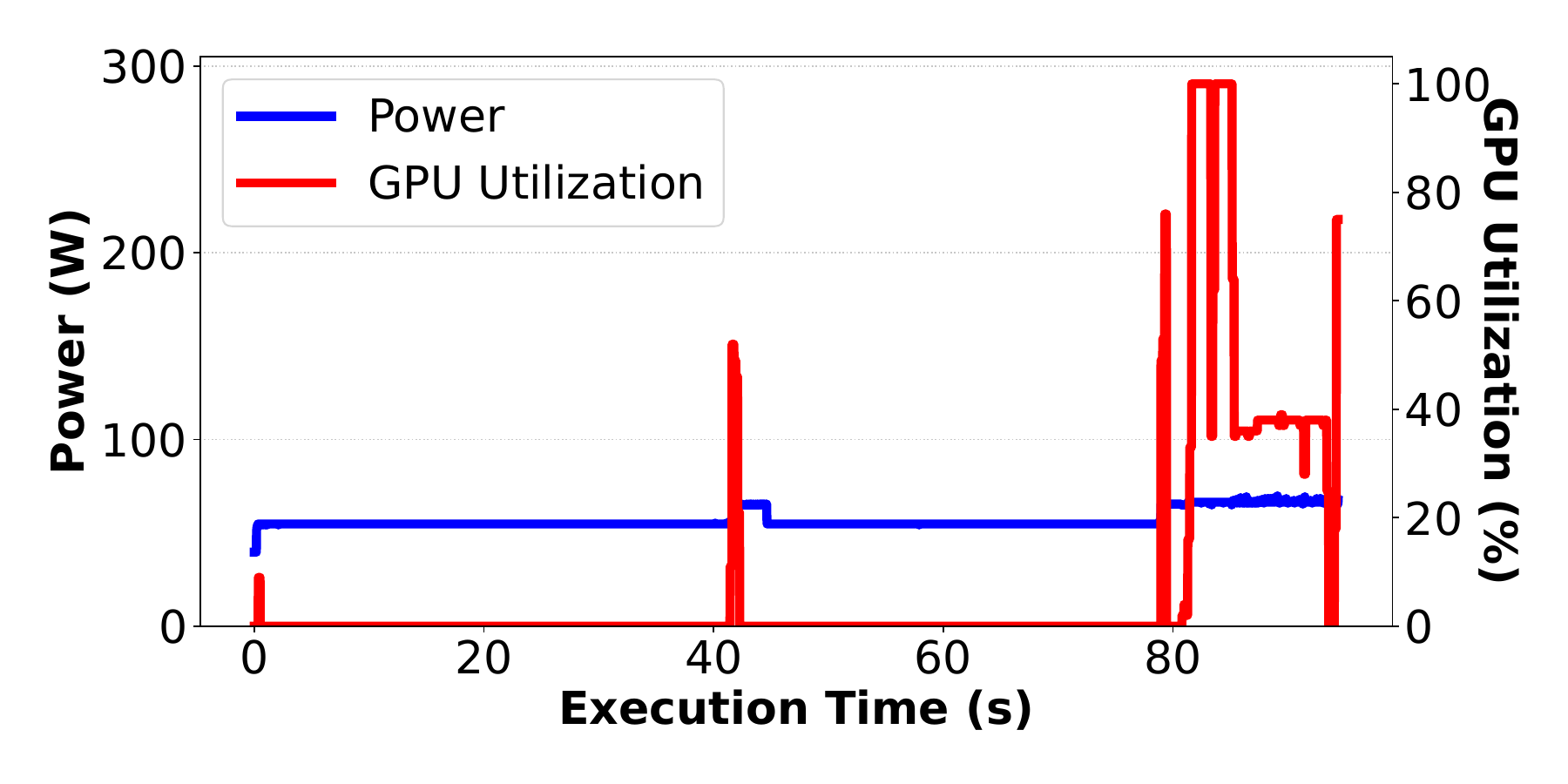}
        \caption{Power trace from running mixed-precision QMCPACK after removing computations on an air-cooled V100 GPU.} %
        \label{fig:qmcpack-mix-new}
        \end{subfigure}
    \caption{Comparing power traces on QMCPACK before and after changing the application not to perform unnecessary computations.}
    \Description{Stacked graph of different types of predictions compared to the ground truth for different benchmarks.}
    \label{fig:qmcpack-observation}
    \vspace{-2ex}
\end{figure}

\begin{figure}
    \centering
    \includegraphics[scale=0.19]{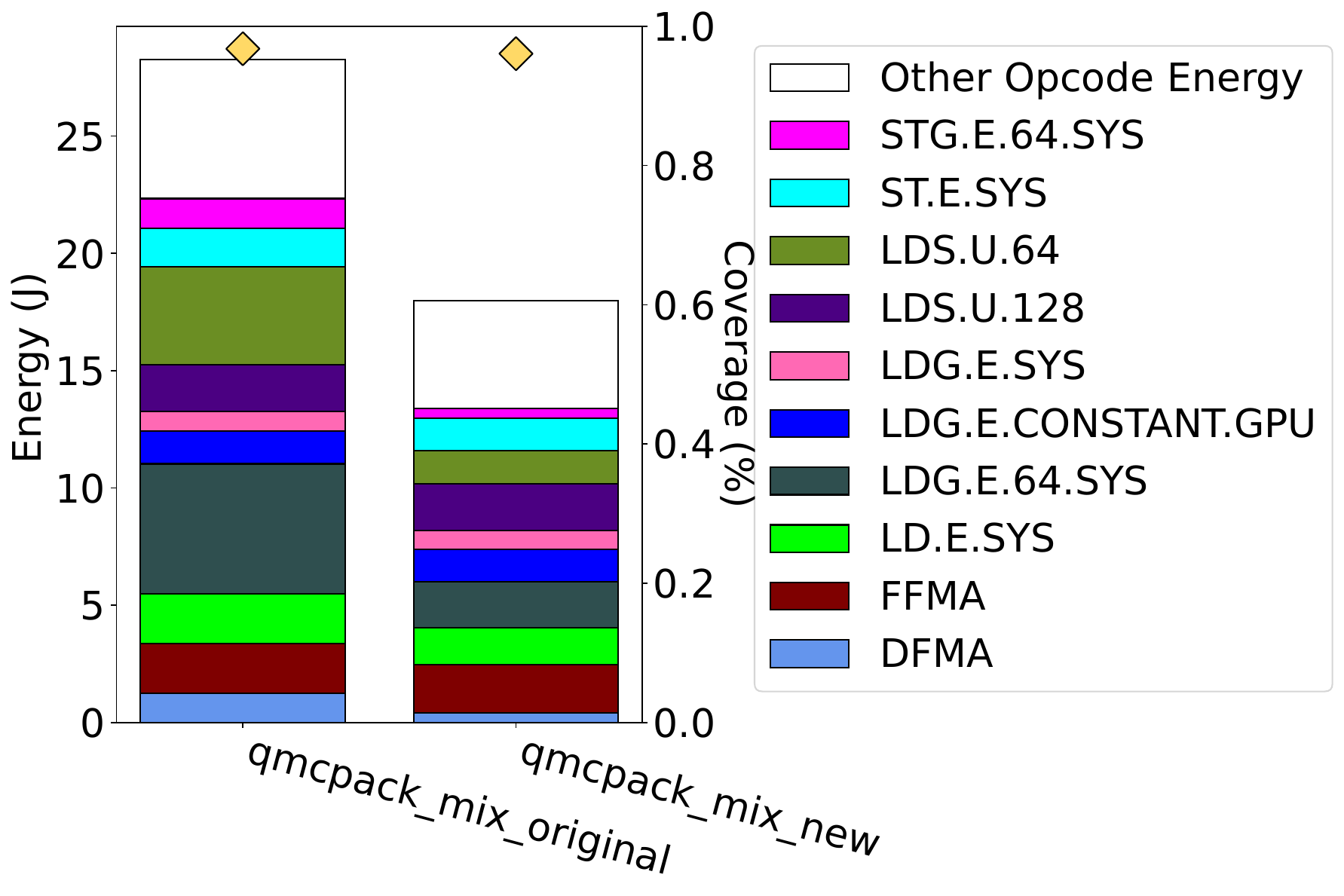}
    \caption{Energy prediction before and after the updating mixed-precision QMCPACK.}
    \Description{Scatterplot comparing.}
    \label{fig:qmcpack-prediction}
    \vspace{-2ex}
\end{figure}

To study QMCPACK, we collaborated with the QMCPACK developers and integrated \DESIGN{} into 
a monitoring workflow at ORNL.
When monitoring mixed-precision QMCPACK runs, \DESIGN{} identified unusual DMC spikes that were absent in full-precision QMCPACK.
By leveraging \DESIGN{}'s fine-grained energy profile, we identified an error in the mixed-precision QMCPACK code that caused the application to call a function at a higher frequency than intended, unintentionally reducing scientific throughput.
Figure~\ref{fig:qmcpack-mix-original} shows the original utilization trace, and Figure~\ref{fig:qmcpack-mix-new} shows the utilization trace after the change -- Figure~\ref{fig:qmcpack-mix-original} has much more prominent red spikes.
Figure~\ref{fig:qmcpack-prediction} shows the impact of this change on one walker over two instances of the update: \DESIGN{} predicts a 36\% reduction in GPU energy consumption, 1\% off the 35\% observed real GPU reduction. %

\section{Discussion}
\label{sec:discussion}

\noindent
\textbf{Limitations}:
\textit{SM activity}: When training the model with microbenchmarks, we run the GPU with all SMs active.
This amortizes the shared resource cost across all SMs and ensures that the energy spent is primarily on induced activity rather than dominated by the static or constant energy of idling SMs.
However, in full-sized applications, it is not guaranteed that all SMs are active.
To match the activity, a full sweep of activity levels could improve our accuracy.
Nevertheless, providing all combinations of SM activity and instruction mixes leads to an intractable design-space explosion, so we focus on the typical case.

\noindent
\textit{Deep Instruction Pipelines -- Hiding and Attributing Power}:
\DESIGN{} splits energy attribution into constant, static, and individual instructions.
Since instruction pipelines are complex, this attribution strategy is not perfect. 
Prior work showed that direct energy measurements of very simple pipelined architectures are possible~\cite{Chang2000}. 
However, in complex GPUs, instructions cannot be fully isolated, occupancy and dark silicon within a pipeline make a difference~\cite{Huerta2025}, and latency hiding is a big part of performance~\cite{Lee2014}.
Thus, a clear attribution is challenging.
Nevertheless, quantifying and accounting for this error could further improve \DESIGN{}.

\noindent
\textit{Measurement Granularity}
We use NVML as the ground truth for our energy measurements.
However, NVML is known to have a coarse measurement granularity~\cite{Yang2024-parttimeNvidiasmi}.
As kernels can reach sub-millisecond durations, actually measuring the energy consumed by these kernels may not be accurate.
While we can predict using high-fidelity instruction profiling, comparisons with the ground-truth energy measurements can be affected by limitations of publicly available measurement tools.

\noindent
\textit{CUDA, PTX, \& SASS Compiler Optimizations}:
We use CUDA version 11.0 for V100 GPUs, and CUDA version 12.0 for A100 and H100 GPUs (Section~\ref{subsec:meth-apps}).
While there would be no impact when predicting applications compiled on different CUDA versions, silent compiler changes can affect training \DESIGN{}.
In particular, microbenchmarks that produce particular instruction sequences may not make the same sequence if compiled with a different CUDA version.
Although CUDA allows inlined assembly, 
the assembler still decides which SASS instructions to use

\noindent
\textit{Collective communication} \DESIGN{} currently only models the instructions for a single GPU.
However, since HPC applications typically use multiple GPUs, \DESIGN{} currently does not model the inter-GPU communication in these workloads.
In subsequent work, we are working on extending \DESIGN{} to model these additional features~\cite{TranMaiterth2026-et}.

\begin{figure}[tb!]
    \centering
    \includegraphics[width=0.85\linewidth]{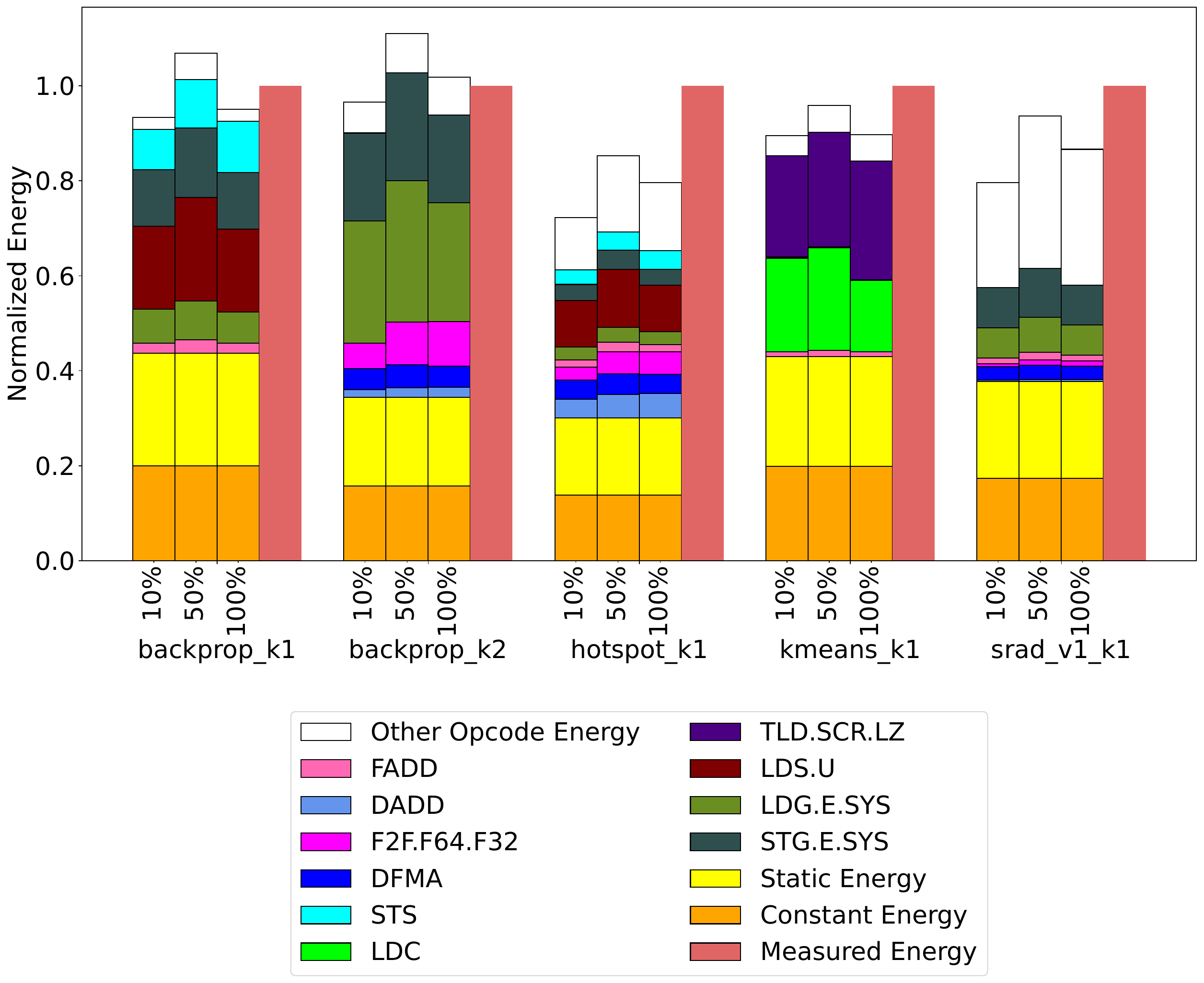}
    \caption{Applying \DESIGN{} while using a fraction (10\% and 50\%) of Summit per-instruction energy data as input and applying an affine transform to obtain the remaining instruction energies. The 10\% fraction, 50\% fraction, and full table achieve MAPEs of 13\%, 10\%, and 14\%, respectively.}
    \Description{Applying \DESIGN{} while using a fraction (10\% and 50\%) of Summit per-instruction energy data as input and applying an affine transform to obtain the remaining instruction energies. The 10\% fraction, 50\% fraction, and full table achieve MAPEs of 13\%, 10\%, and 14\%, respectively.}
    \label{fig:predict_summit_with_cloudlab}
    \vspace{-2ex}
\end{figure}
\noindent
\textbf{Profiler Overhead}:
As the microbenchmark suite takes in iteration count as input, the duration is tunable.
To profile energy, we ran five repetitions per microbenchmark, each lasting 180 seconds.
We allow for an additional 60 seconds after the run completes for the GPU to cool down.
To profile instruction counts and caching behavior, as the microbenchmarks are well-defined, and instruction counts scale with the iteration count. We can run the profiler for fewer iterations and scale up to the appropriate value. %
For instruction counts, we scale them to match the iteration count used in the energy calculation.
For caching behavior, we use the hit rates from the smaller iteration count.
Profiling for instructions and caching for each GPU architecture took less than two hours.
However, this is a one-time cost per GPU.

Since profiling can be time-consuming, we also examine the relationship between the per-instruction energy tables across different systems.
Looking at the per-instruction energy tables for the air-cooled V100 and water-cooled V100, we observe a strong linear relationship with an $R^2$ of $0.988$.
We leverage this relationship to reduce the number of microbenchmarks to profile.
Specifically, we begin by selecting a random subset of the per-instruction energy tables from both systems to fit a linear regression.
We then predict the remaining per-instruction energies for the water-cooled system using a linear regression on the air-cooled system energy table.
The results in Figure~\ref{fig:predict_summit_with_cloudlab} show that using a random subset of 10\% of the instructions gives us 13\% MAPE,  while using 50\% of the instructions gives 10\% MAPE, and using all of the gathered data from the water-cooled system provides 14\% MAPE.
Thus, a linear relation can accelerate the construction of the energy table and maintain the same level of accuracy.

\noindent
\textbf{Applicability to Other GPU Vendors}: We evaluated \DESIGN{} on widely used NVIDIA GPUs.
However, the US DOE's recent, GPU-rich Aurora, El Capitan, and Frontier supercomputers utilize GPUs from other vendors.
\DESIGN{}'s methodology can be adapted to other vendor ISAs.
Broadly, \DESIGN{} requires three inputs: instruction counts, power measurements, and cache hit rates.
Assuming any architecture has tools that provide this information, one can construct \DESIGN{}'s per-instruction energy model.
For example, since AMD GPUs do not have an intermediate ISA (Section~\ref{subsec:back-gpuInstr}), they generate assembly code for specific AMD GPUs.
Thus, it will be simpler to microbenchmark specific AMD GPU assembly instructions~\cite{GutierrezBeckmann2018-amdGem5, JamiesonChandrashekar2022-gap}.

\noindent
\textbf{Applicability to Other NVIDIA GPUs}: We examine \DESIGN{} on Volta-, Ampere-, and Hopper-class GPUs.
However, NVIDIA has released newer GPUs.
Although ISA and microarchitecture differ across generations (e.g., enhanced support for sparse tensor operations), NVIDIA's virtual PTX ISA makes applying \DESIGN{} to newer NVIDIA GPUs relatively straightforward by handling the conversion to architecture-specific SASS instructions.

\noindent
\textbf{Alternative Cooling Strategies}: \DESIGN{} can also easily adapt to additional cooling approaches.
Unlike AccelWattch, our measurement methodology (Section~\ref{subsec:des-measure}) focuses on steady state behavior.
Thus, while different cooling approaches (e.g., mineral oil) %
will affect the temperature at which a given GPU runs.
By focusing on the steady state, we ensure that these differences will not affect \DESIGN{}'s overall behavior.

\noindent
\textbf{Simulator Integration}: Although not a focus of our work, the energy per instruction information generated by \DESIGN{} can be integrated with simulators like AccelSim~\cite{LewShah2019-gpgpusimML, KhairyShen2021-accelSim} and gem5~\cite{GutierrezBeckmann2018-amdGem5, LowePowerAhmad2020-gem520}.
For example, \DESIGN{} could replace or augment the AccelWattch model within AccelSim or plug into gem5's new power model interface~\cite{SmithBruce2024-gem5Power, SmithSinclair2025-gem5Power}. %

\section{Related Work}
\label{sec:relWork}

\noindent
\textbf{Energy Efficiency:}
Prior work has examined improving energy efficiency from other perspectives, including site infrastructure, by introducing new cooling solutions~\cite{Acun-hipc17-proactiveCooling} or looking at large-scale modeling~\cite{ljungdhal2022-modelcooling,maiterth2025-digitaltwin}.
Prior work hardware advancements have also developed more accurate power measurement capabilities~\cite{David2010}, sensor-based native measurements from power meters in various locations of power distribution systems~\cite{choi_roofline_2013, shin_revealing_2021}, and native GPU power measurement capabilities~\cite{Yang2024-parttimeNvidiasmi} exposed via proprietary interfaces~\cite{noauthor_rocm_nodate-1, noauthor_rocm_nodate, noauthor_dcgm_nodate, noauthor_nvidia_nodate} or system level in-band or out-of-band interfaces~\cite{bartolini2018davide, noauthor_redfish_nodate, noauthor_pm_nodate, Thaler:2020, openbmc_event}.
In concert with downstream data plumbing and management software~\cite{agelastos_lightweight_2014, bartolini_paving_2019, borghesi_examon-x_2023, eastep_global_2017, netti_dcdb_2020}, these GPU-level mechanisms have helped researchers and systems operators improve GPU energy efficiency in HPC systems.
However, these mechanisms fall short when profiling application energy.
GPU kernel activities frequently occur at the microsecond level, but prior work on profiling GPU kernels in HPC systems profiles second-scale information~\cite{govind_comparing_2023, shin_revealing_2021}.
Thus, these tools struggle to provide practical application analysis.
Furthermore, increasing GPU package size pressures HPC sites, and applications are resorting to multiplexing workloads on sub-GPU constructs such as MIG~\cite{nvidia_mig}, MxGPU~\cite{mxgpu}, or chiplets in multi-chiplet GPUs~\cite{amd_mi200_gcds, Choquette2023-hopper, DalmiaShashiKumar2024-cpelide, LohSchulte2023-mi250, SmithLoh2024-mi300A}.
This further complicates the attribution of energy to applications without requiring additional hardware sensors.
Consequently, measurements at the required granularity are not supported~\cite{Yang2024-parttimeNvidiasmi}.
Other work has employed software management techniques to reduce GPU energy consumption while maximizing performance.
This includes determining optimal GPU frequency for running workloads~\cite{fan2019, schoonhoven2022, you2023-zeus} and developing energy- and power-aware schedulers~\cite{maiterth_energy_2018, ahn2020-flux,patel2024-polca, stojkovic2025-tapas,stojkovic2025-dynamollm,hanindhito2025}.
Although these works can also lower the energy consumption of GPU applications, they focus on tuning system parameters to improve energy efficiency on a given system.
Conversely, \DESIGN{} enables developers and researchers to modify the workload itself (e.g., our Section~\ref{subsec:eval-casestudy} case studies) to improve energy efficiency.
Thus, while both \DESIGN{} and prior work improve energy efficiency, their respective focuses differ; \DESIGN{} could also be combined with prior work to further enhance efficiency.

\noindent
\textbf{Architecture-level GPU Energy Modeling}:
In addition to the energy modeling tools (Table~\ref{tab:relWork}), the modeling and simulation community also uses other CPU or GPU architecture-level energy modeling approaches.
Broadly, the current energy-aware approaches are: first principle measurements~\cite{Li2013-mcpat, muralimanohar2009cacti, cacti}, projecting empirical energy measurements on existing systems to future systems~\cite{ArafaElWazir2020-nmsu, BrooksTiwari2000-wattch, DelestracMiquel2024-montpellier, KandiahPeverelle2021-accelWattch, LengHetherington2013-gpuWattch}, machine learning models to predict energy consumption~\cite{fan2019,guerreiro2018,guerreiro2019,WuGreathouse2015-amdGPUPowerML}, obtaining specific components' energy estimates from design tape outs, or low-level (e.g., Spice) models to get specific components' energy estimates.
Each of these approaches has issues.
CACTI and McPAT have not been updated in over 8 years and are no longer representative of modern components.
Likewise, design tape outs are time consuming, expensive, and %
often guarded by vendors as proprietary information.
Low-level Spice models are accurate, but scale poorly to increasingly large, complex systems.
Finally, as we show in Section~\ref{sec:intro}, even the state-of-the-art AccelWattch gives inaccurate results under even minor configuration perturbations.
Thus, these techniques are too inaccurate, inflexible, or outdated.
Hence, \DESIGN{} significantly improves GPU energy modeling.

\section{Conclusion}
\label{sec:conclusion}

As HPC systems adopt ever larger numbers of accelerators to meet the exponential computing demands of modern workloads, they become increasingly power-constrained -- a trend likely to continue. %
Thus, it is essential to co-design HPC systems for both energy and performance, but 
existing modeling and simulation tools cannot provide flexible, high-fidelity energy models.
We present \DESIGN{}, a new approach for measuring and modeling GPU energy consumption that overcomes these shortcomings.
Across 16 popular GPGPU, graph analytics, HPC, and ML workloads, \DESIGN{} produces a MAPE of 14\% compared to 32\% from state-of-the-art.
Furthermore, across water-cooled V100 GPUs and more modern architectures like A100 and H100 GPUs, \DESIGN{}'s accuracy remains consistently at or below a MAPE of 15\%.
Finally, our case studies demonstrate how HPC application developers can use \DESIGN{} to optimize their workloads to improve performance of applications on these accelerator-rich clusters given a fixed energy budget.

\begin{acks}
\label{sec:acknowledge}
This research used resources of Oak Ridge Leadership Computing Facility at the Oak Ridge National Laboratory, which is supported by the Office of Science of the U.S. Department of Energy under Contract No. DE-AC05-00OR22725.
This research was supported in part by an appointment to the U.S. Department of Energy's OMNI Technology Alliance Internship Program, sponsored by DOE and administered by the Oak Ridge Institute for Science and Education (ORISE).
This research used resources of the Experimental Computing Laboratory (ExCL) at the Oak Ridge National Laboratory, which is supported by the Office of Science of the U.S. Department of Energy under Contract No. DE-AC05-00OR22725. 
The authors acknowledge the Texas Advanced Computing Center (TACC)~\cite{tacc} at The University of Texas at Austin for providing computational resources that have contributed to the research results reported within this paper and Zhao Zhang in particular for arranging access to the cluster. 
Finally, this work is supported by NSF grant CNS-2312688 and NSF CIRC-2450448.
\end{acks}

\bibliographystyle{ACM-Reference-Format}
\bibliography{references,gpu_power_measure}

\end{document}